\documentclass[twocolumn,amsmath,amssymb,floatfix,prl,footinbib,superscriptaddress]{revtex4-1}

\usepackage{amsmath,amssymb,natbib,bm,graphicx,url,epsf,color}
\usepackage[english]{babel}
\usepackage{wasysym}
\usepackage{MnSymbol}
\usepackage{amsmath}
\usepackage{graphicx}
\usepackage{pdfpages}
\usepackage[colorinlistoftodos]{todonotes}
\usepackage[colorlinks=true, allcolors=blue]{hyperref}
\usepackage[labelfont=bf,justification=raggedright,labelsep = space]{caption}

\usepackage{siunitx}

\newcommand{\QH}{QH~}
\newcommand{\HCB}{HCB~}
\newcommand{\CNP}{CNP~}
\newcommand{\CI}{CI}

\definecolor{wine}{RGB}{127,0,0}
\definecolor{navy}{RGB}{0,0,119}

\newcommand{\Vg}{V_{\mathrm{g}}}

\newcommand{\Jq}{J_{\mathrm{Q}}}

\newcommand{\Jqe}{J_{\mathrm{Q}}^\mathrm{e}}
\newcommand{\EC}{E_{\mathrm{C}}}
\newcommand{\Tc}{T_{\mathrm{c}}}

\newcommand{\DeltaTc}{\Delta T_{\mathrm{c}}}

\newcommand{\kB}{k_{\mathrm{B}}}

\newcommand{\Gtwopt}{G_{2\mathrm{pt}}}

\makeatletter
\AtBeginDocument{\let\LS@rot\@undefined}
\makeatother

\makeatletter
\def\NAT@spacechar{}
\makeatother

\begin{document}
\title{Quantized heat flow in the Hofstadter butterfly}

\author{A. Zhang}
\affiliation{Universit\'e Paris-Saclay, CEA, CNRS, SPEC, 91191 Gif-sur-Yvette cedex, France
}
\author{G. Aissani}
\affiliation{Laboratoire de Physique de l’Ecole normale sup\'erieure, ENS, Universit\'e PSL,
CNRS, Sorbonne Universit\'e, Universit\'e Paris Cit\'e, F-75005 Paris, France
}
\author{Q. Dong}
\affiliation{CryoHEMT, 91400 Orsay, France
}
\author{Y. Jin}
\affiliation{Universit\'e Paris-Saclay, CNRS, Centre de Nanosciences et de Nanotechnologies (C2N), 91120 Palaiseau, France
}

\author{K. Watanabe}
\affiliation{Research Center for Electronic and Optical Materials, National Institute for Materials Science, 1-1 Namiki, Tsukuba 305-0044, Japan
}
\author{T. Taniguchi}
\affiliation{Research Center for Materials Nanoarchitectonics, National Institute for Materials Science,  1-1 Namiki, Tsukuba 305-0044, Japan
}
\author{C. Altimiras}
\affiliation{Universit\'e Paris-Saclay, CEA, CNRS, SPEC, 91191 Gif-sur-Yvette cedex, France
}
\author{P.~Roche}
\affiliation{Universit\'e Paris-Saclay, CEA, CNRS, SPEC, 91191 Gif-sur-Yvette cedex, France
}
\author{J.-M. Berroir}
\affiliation{Laboratoire de Physique de l’Ecole normale sup\'erieure, ENS, Universit\'e PSL,
CNRS, Sorbonne Universit\'e, Universit\'e Paris Cit\'e, F-75005 Paris, France
}
\author{E. Baudin}
\affiliation{Laboratoire de Physique de l’Ecole normale sup\'erieure, ENS, Universit\'e PSL,
CNRS, Sorbonne Universit\'e, Universit\'e Paris Cit\'e, F-75005 Paris, France
}
\author{G. F\`eve}
\affiliation{Laboratoire de Physique de l’Ecole normale sup\'erieure, ENS, Universit\'e PSL,
CNRS, Sorbonne Universit\'e, Universit\'e Paris Cit\'e, F-75005 Paris, France
}
\author{G. M\'enard}
\affiliation{Laboratoire de Physique de l’Ecole normale sup\'erieure, ENS, Universit\'e PSL,
CNRS, Sorbonne Universit\'e, Universit\'e Paris Cit\'e, F-75005 Paris, France
}
\author{O. Maillet}
\affiliation{Universit\'e Paris-Saclay, CEA, CNRS, SPEC, 91191 Gif-sur-Yvette cedex, France
}
\author{F.D. Parmentier}\email[Corresponding author: ]{francois.parmentier@phys.ens.fr}
\affiliation{Universit\'e Paris-Saclay, CEA, CNRS, SPEC, 91191 Gif-sur-Yvette cedex, France
}
\affiliation{Laboratoire de Physique de l’Ecole normale sup\'erieure, ENS, Universit\'e PSL,
CNRS, Sorbonne Universit\'e, Universit\'e Paris Cit\'e, F-75005 Paris, France
}

\date{\today}

\maketitle

\textbf{When subjected to a strong magnetic field, electrons on a two-dimensional lattice acquire a fractal energy spectrum called Hofstadter's butterfly. In addition to its unique recursive structure, the Hofstadter butterfly is intimately linked to non-trivial topological orders, hosting a cascade of ground states characterized by non-zero topological invariants. These states, called Chern insulators, are usually understood as replicas of the ground states of the quantum Hall effect, with electrical and thermal conductances that should be quantized, reflecting their topological order. The Hofstadter butterfly is now commonly observed in van-der-Waals heterostructures-based moiré superlattices. However, its thermal properties, particularly the quantized heat flow expected in the Chern insulators, have not been investigated, potentially questioning their similarity with standard quantum Hall states. Here we probe the heat transport properties of the Hofstadter butterfly, obtained in a graphene~/~hexagonal boron nitride moiré superlattice. We observe a quantized heat flow, uniquely set by the topological invariant, for all investigated states of the Hofstadter butterfly: quantum Hall states, Chern insulators, and even symmetry-broken Chern insulators emerging from strong electronic interactions. Our work firmly establishes the universality of the quantization of heat transport and its intimate link with topology.}

\section{Main}

\begin{figure*}[ht!]
\centering
\includegraphics[width=0.98\textwidth]{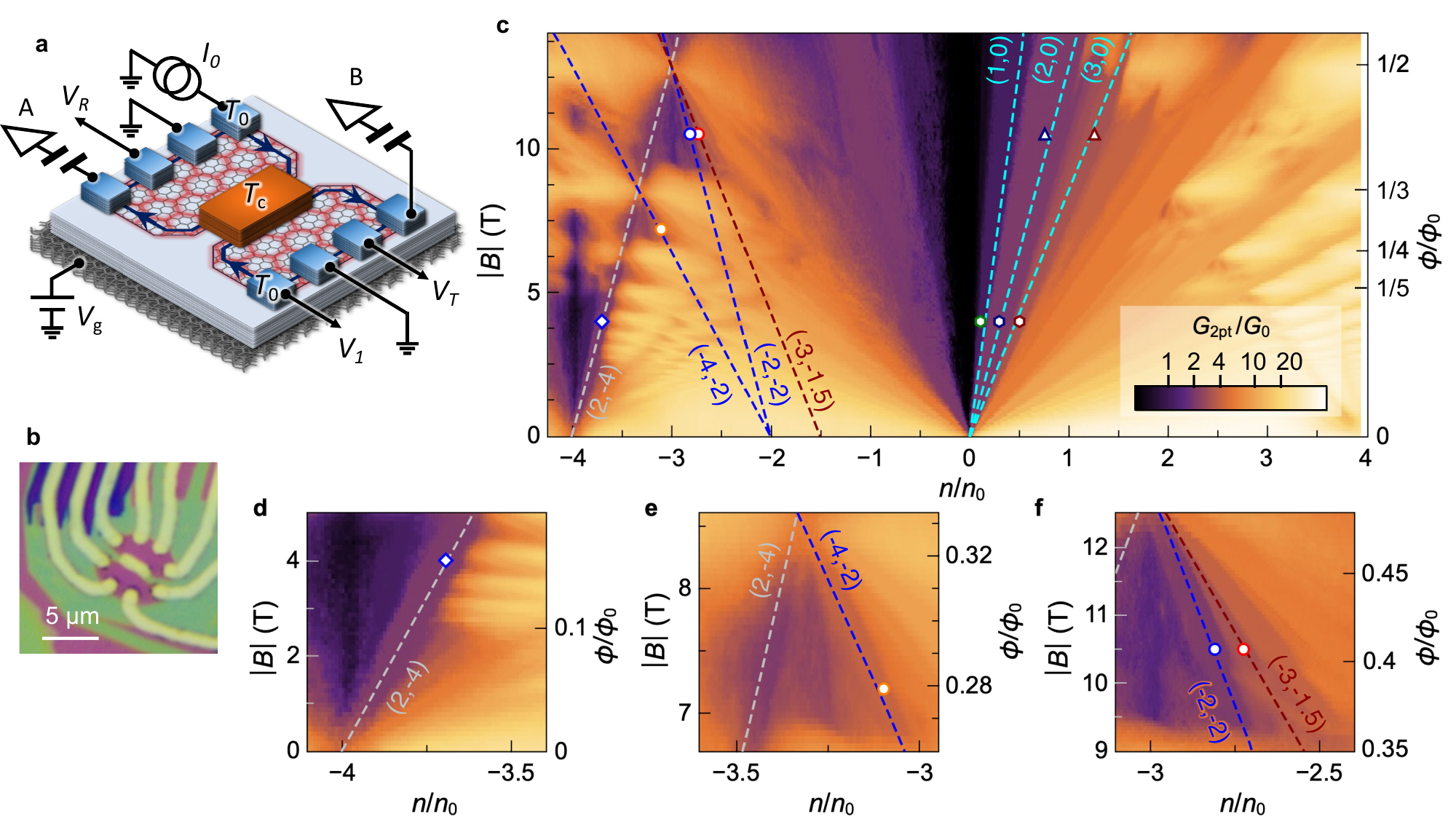}
-\caption{\label{fig1} \textbf{$\vert$ Hofstadter spectrum in a graphene/hBN moiré heterostructure. a,} Sample schematic and experimental wiring. \textbf{b,} Optical micrograph of the sample. \textbf{c,} 2-point differential conductance $(dV_R/dI_R)^{-1}$ (in log scale) versus moiré unit cell filling $n/n_0$ and magnetic field $\lvert B\rvert$ (left Y-axis) / reduced magnetic flux $\phi/\phi_0$, measured at $T\approx20~$mK. Dashed lines indicate the probed \QH and Chern insulators states, along with their corresponding $(t,s)$. The line color correspond to the value of $s$. Symbols indicate where we performed the heat transport measurements shown here: \textcolor{blue}{\Large $\diamond$}: (2,-4) at $B=+4~$T; \textcolor{orange}{\Large$\circ$}: (-4,-2) at $B=-7.2~$T; \textcolor{blue}{\Large$\circ$}: (-2,-2) at $B=-10.5~$T;  \textcolor{red}{\Large$\circ$}: (-3,-1.5) at $B=-10.5~$T;  \textcolor{olive}{\textbf{$\varhexagon$}}:~(1,0) at $B=+4~$T;   \textcolor{navy}{\textbf{$\varhexagon$}}:~(2,0) at $B=+4~$T;   \textcolor{wine}{\textbf{$\varhexagon$}}:~(3,0) at $B=+4~$T;   \textcolor{navy}{$\medtriangleup$}:~(2,0) at $B=+10.5~$T;   \textcolor{wine}{$\medtriangleup$}:~(3,0) at $B=+10.5~$T. \textbf{d, e,} and \textbf{f,} zoom of the data shown in \textbf{c} in the vicinity of the satellite peak (\textbf{d}) and of the \CNP clones at $\phi/\phi_0=1/3$ (\textbf{e}) and $\phi/\phi_0=1/2$ (\textbf{f}).} 
\end{figure*}

The concept of topology in condensed matter physics is embodied by the quantum Hall (\QH) effect~\cite{Thouless1982}, occurring in a two-dimensional (2D) electron system immersed in a strong perpendicular magnetic field $B$. The quantization of the electronic spectrum into discrete Landau levels gives rise to gapped states with a non-trivial topology manifesting itself through chiral edge states with quantized transport properties. The electrical conductance $G$ is quantized in units of the electrical conductance quantum $G_0=e^2/h$ ($e$ is the electron charge and $h$ Planck's constant), and the thermal conductance in units of the thermal conductance quantum $\kappa_0 T= \frac{\pi^2\kB^2}{3h}T$~\cite{Kane1997,Jezouin2013a} ($\kB$ is Boltzmann constant, and $T$ the temperature). Crucially, in the integer \QH effect, the quantization prefactor for both conductances is the same integer number, given by the filling factor $\nu=n h/eB$, where $n$ is the 2D charge carrier density. 

In presence of a periodic lattice~\cite{Harper1955}, the spectrum changes radically, showing a fractal series of gapped states with non-trivial topology, referred to as Hofstadter's butterfly~\cite{Hofstadter1976}. This can be observed in graphene-based moiré heterostructures, for instance in graphene crystallographically aligned with hexagonal boron nitride (hBN)~\cite{Dean2013,Hunt2013,Ponomarenko2013,Wang2015,Yang2016,Spanton2018}, or in twisted bilayer graphene~\cite{Nuckolls2020,Das2021,Pierce2021,Saito2021,Wu2021,Xie2021,Yang2022,Yu2022,He2025}. In these systems, the moiré potential plays the role of the periodic lattice with a constant in the $\lambda\sim10~$nm range, allowing the observation of Hofstadter's butterfly at reasonable magnetic fields ($5-40~$T). The topological states appear as lines in the plane defined by the magnetic flux per moiré unit cell in units of the magnetic flux quantum $\phi/\phi_0=(B\times\sqrt{3}\lambda^2/2)/(h/e)$ and the number of electrons per moiré unit cell $n/n_0=n\sqrt{3}\lambda^2/2$~\cite{Hunt2013}, parametrized by the diophantine equation:
\begin{equation}
n/n_0=t\times \phi/\phi_0+s.
\label{eq:diophantine}
\end{equation}

The numbers $(t,s)$ are integer or fractional, and uniquely define the topological states. In particular, the slope $t$ sets the topological invariant (or Chern number) of the gaps~\cite{Thouless1982} as well as the quantized edge electrical conductance through the Streda formula $G=t\times G_0$~\cite{Streda1982}. The various topological states of the Hofstadter butterfly are classified according to the value of $t$ and $s$. Integer (respectively fractional) \QH states have $s=0$ and integer (respectively fractional) $t$ (which then identifies to the filling factor $\nu$). States with $s=\pm4$ can also be classified as \QH since they correspond to a full filling of the moiré unit cell, taking into account the spin and valley symmetries. \textit{Chern insulators} (CIs) have integer $t$ and $s$, and correspond to the fractal gaps of the Hofstadter butterfly in a single-particle picture. Strong electronic interactions then give rise to additional topological states: \textit{symmetry-broken Chern insulators} (SBCIs) have integer $t$ but fractional $s=p/q$, and are associated to a breaking of the translational symmetry of the moiré superlattice where the unit cell is spontaneously enlarged to $q$ times the moiré unit cell~\cite{Wang2015,Spanton2018}. Finally, fractional Chern insulators (FCIs)~\cite{Spanton2018}, which have both fractional $t$ and $s$, are analogous to fractional \QH states, corresponding to the stabilization of new correlated ground states at fractional filling of the bands of the Hofstadter butterfly.

Quantized electrical conductances were measured in \CI s and SBCIs~\cite{Hunt2013,Wang2015,Das2021,Saito2021,Wu2021,He2025}, as well as more recently in the zero-field FCIs emerging in twisted MoTe$_2$~\cite{Park2023} and rhombohedral graphene~\cite{Lu2024} (but not, to our knowledge, in the finite field FCIs of graphene/hBN moirés and twisted bilayer graphene). However, the quantization of heat transport remains to be demonstrated for the topological states of the Hofstadter butterfly. This has important implications on the generalization to heat transport of the Streda formula, as well as on the understanding of \CI s through their analogy with \QH states, where heat flow is quantized for both integer~\cite{Jezouin2013a,Srivastav2019} and fractional~\cite{Banerjee2017,Banerjee2018,Srivastav2019,LeBreton2022} filling factors. In particular, heat transport can probe the presence or absence of heat-carrying neutral modes, either along the edges as in the fractional \QH effect~\cite{Kane1994,Banerjee2017,LeBreton2022}, or in the bulk, due to spontaneous symmetry breakings~\cite{Pientka2017,Delagrange2024}. Thus, heat transport is in a sense an even stronger signifier of topology, as a quantized heat flow necessarily implies a quantized electrical conductance, while the opposite is not always true.

\section{Heat transport geometry}

Fig.~\ref{fig1}a shows a schematic of our experiment, where we adapt the heat transport technique developed for \QH systems~\cite{Jezouin2013a,Srivastav2019,LeBreton2022} to a graphene / hBN moiré heterostructure. This geometry is based on a small ($1~$\unit{\um}$~\times~3~$\unit{\um}) central metallic island, shown as a red brick in Fig.~\ref{fig1}a, which is heated up to the electron temperature $\Tc$ by a constant Joule power $\Jq$. The island exchanges heat with colder electrodes (blue bricks in Fig.~\ref{fig1}a), kept at base electron temperature $T_0$, through two parts of the graphene flake on either side of the island. Assuming that charge and heat are only carried by the $N$ ballistic channels flowing along the edges of the sample, the heat balance reads:

\begin{equation}
\Jq=2N\Jqe=2N\frac{\kappa_0}{2}(\Tc^2-T_0^2),
\label{eq:heatbal}
\end{equation}

where $\Jqe=\frac{\kappa_0}{2}(\Tc^2-T_0^2)$ is the universal quantum limit of heat flow~\cite{Pendry1983,Jezouin2013a}. We probe Eq.~\ref{eq:heatbal} by connecting the cold electrodes to different specific lines of a dilution refrigerator, according to the chirality of current flow indicated by the blue arrows in Fig.~\ref{fig1}a. A dc current $I_0$ is applied to the current feed electrode (topmost blue brick in Fig.~\ref{fig1}a) to generate the heating power $\Jq=I_0^2/(4NG_0)$~\cite{Jezouin2013a}. Following clockwise on the top half of the device, the thermal noise of the sample (yielding the temperature $\Tc$) is measured downstream of the island through a noise measurement lines (labeled A), and the electrical conductance is measured through the next electrode through its voltage drop $V_R$. Finally, the electrode between the conductance measurement and current feed electrodes is connected to a cold ground. The wiring configuration is symmetric on the other side of the sample, with a noise measurement electrode (B), a conductance measurement electrode (voltage drop $V_T$), and a cold ground. The bottommost current feed electrode (labeled $V_1$) is left floating in the measurements shown here. 

In presence of chiral edge states (blue lines in Fig.~\ref{fig1}a) the current $I_0$ applied on the current feed electrode flows to the metallic island, which splits it into its two outgoing edge states, similarly to a node in an electrical circuit. The split currents then flow to the cold grounds on either side of the sample, and we detect the fluctuations and average value of the voltage drop developing on (respectively) the noise and conductance measurement electrodes. Conductance measurements on either side are labeled with respect to the $I_0$ current feed electrode shown in Fig.~\ref{fig1}a: \textit{T} for transmitted across the metallic island, \textit{R} for reflected at the metallic island. The 2-point conductance $\Gtwopt=(dV_{R/T}/dI_{R/T})^{-1}$, obtained by measuring the ac voltage response of one electrode to a small ac current $dI_{R/T}$ (typically 40~pA) applied to the same electrode, reflects the Hall conductance of the device~\cite{Hunt2013,LeBreton2022}. 

The device, shown in Fig.~\ref{fig1}b, is based on a monolayer graphene flake encapsulated between two 50~nm-thick hBN crystals, one of which is aligned with the graphene flake (see Methods section for details on the alignment and encapsulation procedure). The heterostructure includes a global graphite back-gate upon which we apply a dc voltage $\Vg$ to tune the carrier type and density, and the metallic electrodes (including the central island) are made using the edge contacts technique~\cite{Wang2013}. 

\section{Topological states in a graphene/\lowercase{h}BN Hofstadter butterfly}

\begin{figure}[ht]
\centering
\includegraphics[width=0.49\textwidth]{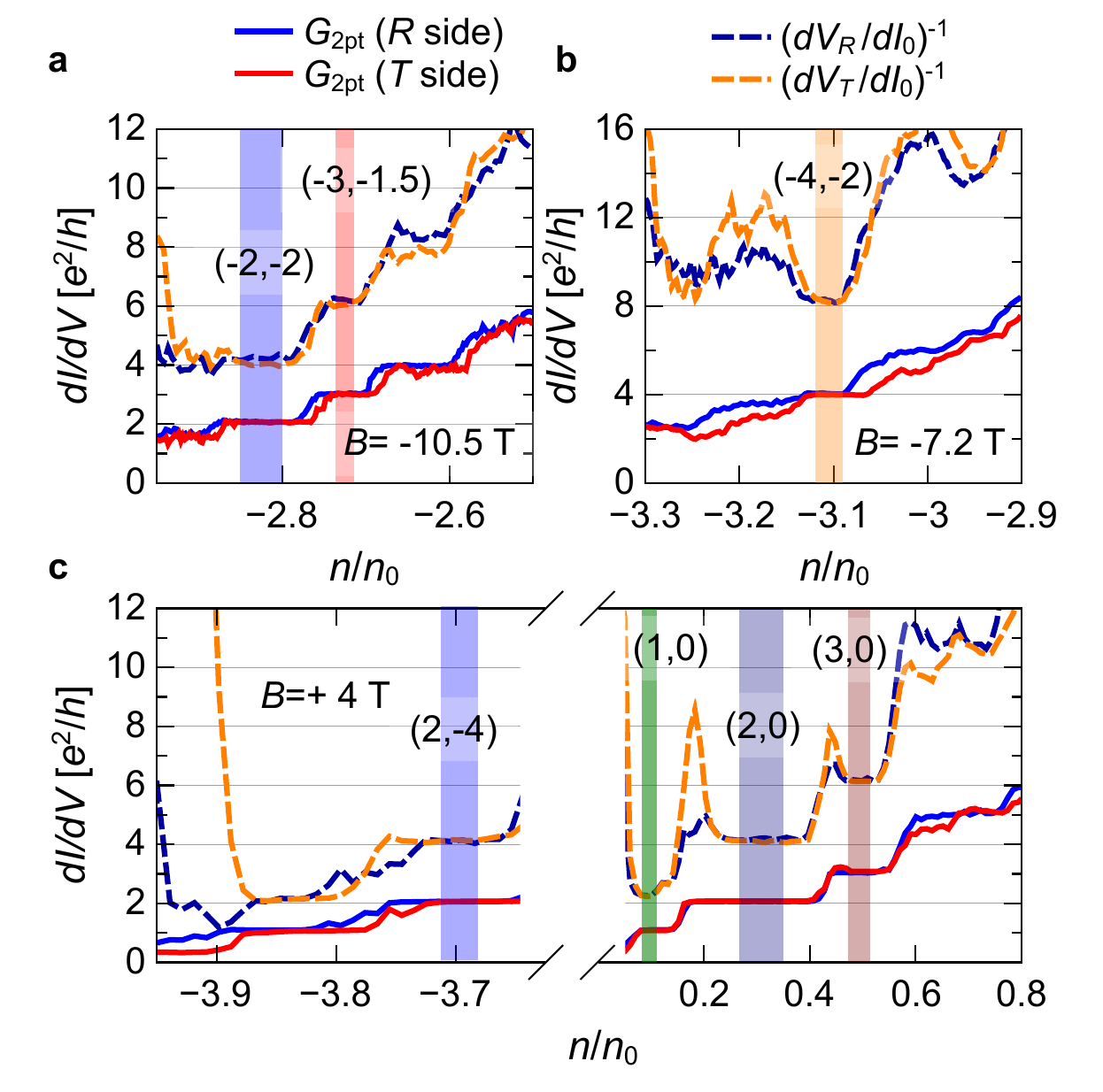}
\caption{\label{fig2} \textbf{$\vert$ Chiral electronic current splitting in Chern and \QH insulators. } Measured differential conductances versus $n/n_0$ at $B=-10.5~$T (\textbf{a}), $B=-7.2~$T (\textbf{b}), and $B=+4~$T (\textbf{c}), for $T=20~$mK. Full lines: 2-point conductances $(dV_R/dI_R)^{-1}$ (blue) and $(dV_T/dI_T)^{-1}$ (red). Dashed lines: reflected (dark blue) and transmitted (orange) transconductances $(dV_{R,T}/dI_0)^{-1}$. Shaded areas indicate the \QH and Chern insulators states probed in our experiment, with colors corresponding to the symbols of Figs.~1,~3, and~4.
}
\end{figure}

\begin{figure*}[ht!]
\centering
\includegraphics[width=0.99\textwidth]{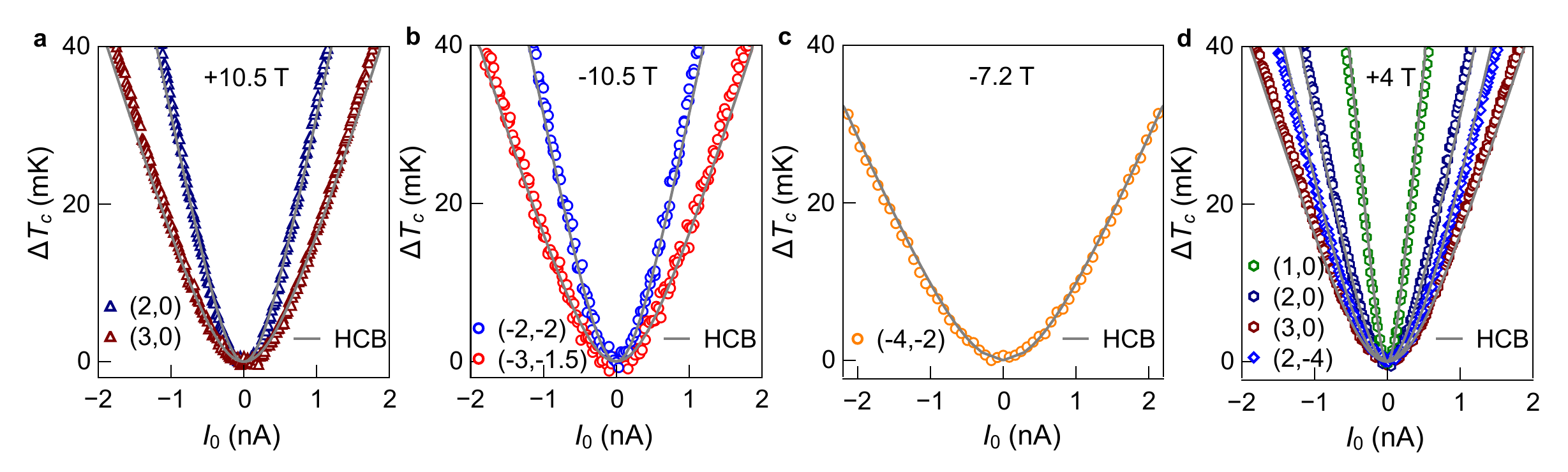}
\caption{\label{fig3} \textbf{$\vert$ Electron thermometry in Hofstadter states.} Measured $\DeltaTc$ versus dc heating current $I_0$, at $B=+10.5~$T (\textbf{a}), $-10.5~$T (\textbf{b}), $-7.2~$T (\textbf{c}), and $+4~$T (\textbf{d}). Symbols: experimental data (shape and color corresponding to the symbols of Fig.~1), taken at $T=20~$mK. Lines: HCB predictions (see text).} 
\end{figure*}

Fig.~\ref{fig1}c shows the dependence of $\Gtwopt$ on the reflected (\textit{R}) side as a function of the applied perpendicular magnetic field $B$ and of the $\Vg$-tuned carrier density, expressed as the filling of the moiré unit cell $n/n_0$. The latter is set by the position of the satellite peaks~\cite{Dean2013,Hunt2013,Ponomarenko2013,SM} in the $B=0~$T measurement of $\Gtwopt(\Vg)$, giving an alignment angle between graphene and hBN $\theta=0.5\pm0.1~^\circ$ and a moiré superlattice constant $\lambda=13.0\pm0.5~$nm. This allows computing the reduced flux $\phi/\phi_0$, shown in the right Y-axis. At finite magnetic field, $\Gtwopt$ displays the hallmark features of the Hofstadter spectrum: Landau fans stemming from the charge neutrality point (CNP) ($s=0$) and both satellite peaks ($s=\pm4$), as well as a fractal set of clones of the CNP arising at fractional values $\phi/\phi_0=1/q$. The CNP clones give rise to various additional fans which correspond to the fractal bands of the Hofstadter butterfly~\cite{Dean2013,Hunt2013,Ponomarenko2013,Spanton2018}, in particular at $\phi/\phi_0=1/3$ and $1/2$. These fans, as well as the CNP and hole satellite peak Landau fans, show conductance plateaus matched by the dashed lines given by the diophantine equation: $(t,s)=(\left\{1,2,3\right\},0)$ (cyan) correspond to \QH insulator states stemming from the CNP. $(t,s)=(2,-4)$ (light grey) corresponds to the \QH insulator state stemming from the hole satellite peak. $(t,s)=(-4,-2)$ and $(t,s)=(-2,-2)$ (blue) correspond to \CI s respectively stemming from the CNP clones at $\phi/\phi_0=1/3$ and $1/2$, and intersecting at half-filling of the moiré unit cell. Finally, $(-3,-1.5)$ (red) corresponds to a symmetry-broken \CI~\cite{Wang2015,Spanton2018} stemming from the CNP clone at $\phi/\phi_0=1/2$. Zooms on the regions near the hole satellite peak (Fig.~\ref{fig1}d) and the CNP clones at $\phi/\phi_0=1/3$ (Fig.~\ref{fig1}e) and $1/2$ (Fig.~\ref{fig1}f) show the extent of the conductance plateau in the $B$, $n/n_0$ plane. 

\section{Chiral charge transport in Chern insulators}

We focus on the charge transport properties of the above \QH and \CI~ states at positions in the magnetic field / density plane indicated by the symbols in Fig.~\ref{fig1}c-f. The plateaus shown in Fig.~\ref{fig1}c-f correspond to a quantized 2-point conductance $t\times G_0$ for all considered states, as shown in Fig.~\ref{fig2}. Importantly, we measure the 2-point conductance on both \textit{R} and \textit{T} sides of the samples (respectively, blue and red lines in Fig.~\ref{fig2}) which show well-quantized $t\times G_0$ plateaus at matching values of $n/n_0$ for all our considered states (see also the Supplementary Information~\cite{SM} for measurements of $\Gtwopt$ on the \textit{T} side). 

Next, we check that the metallic island properly splits the edge current evenly between the two sides of the device. For this we adapt the sign of the applied magnetic field according to that of the topological invariant $t$ of the considered states, so as to obtain the chirality depicted in Fig.~\ref{fig1}a. In addition to the 2-point conductances, we plot in Fig.~\ref{fig2} the transconductances between the current feed electrode (upon which we apply a small ac current $dI_0$ at a different frequency from that of $dI_T$ and $dI_R$, see Methods) and, respectively, the \textit{R} and \textit{T} electrode $(dV_{R}/dI_0)^{-1}$ and $(dV_{T}/dI_0)^{-1}$ (dark blue and orange dashed lines). When the whole sample is in the same topological state with the correct chirality, the transconductances should be equal and quantized to twice the value of the 2-point conductance (see Methods). In the opposite chirality, the transconductances vanish (see Supplementary Information~\cite{SM}), as all the current $I_0$ applied to the sample directly flows to the cold ground, demonstrating the absence of bulk electrical conduction. 

Fig.~\ref{fig2}a and b show the measurements at negative magnetic field ($B=-10.5~$T for Fig.~\ref{fig2}a and $B=-7.2~$T for Fig.~\ref{fig2}b) for the states $(-2,-2)$, $(-3,-1.5)$ and $(-4,-2)$ (blue, red, and orange circles in Fig.~\ref{fig1}). Crucially, in the color shaded regions, both 2-point conductances are quantized and equal at $\Gtwopt=t\times G_0$, while the transconductances are also equal and quantized to $(dV_{R}/dI_0)^{-1}=(dV_{T}/dI_0)^{-1}=2t\times G_0$. This firmly establishes the presence of edge states with the expected chirality and current splitting for both \CI s and SBCI.

Fig.~\ref{fig2}c shows measurements at positive field, $B=+4~$T (data at $+10.5~$T are shown in the Supplementary Information~\cite{SM}), again showing the expected plateaus for the $(2,-4)$ state stemming from the hole satellite peak, as well as the $(1,0)$, $(2,0)$ and $(3,0)$ \QH states stemming from the CNP.

\begin{figure}[ht]
\centering
\includegraphics[width=0.41\textwidth]{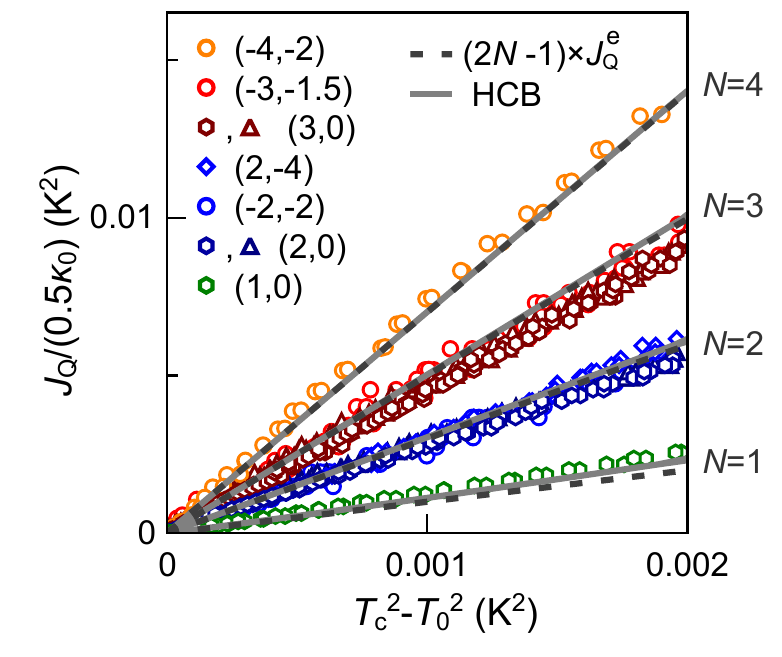}
\caption{\label{fig4} \textbf{$\vert$ Universal quantized heat flow.} Heat flow $\Jq/(0.5 \kappa_0)$ versus $\Tc^2-T_0^2$. Symbols correspond to the data shown in Fig.~3. Dashed lines: quantized heat flow with full suppression of a single heat channel $(2N-1)\Jqe$, with $N=1\rightarrow4$ (from bottom to top) the number of ballistic heat channels on each side of the metallic island. Full lines: \HCB predictions for the corresponding $N$ at $T_0=20~$mK.
}
\end{figure}  

\section{Quantized thermal transport}

The demonstration of the proper chirality and current splitting for all states allow us to formally identify $t$ as the number of ballistic charge-carrying channels. Thus, the Joule power dissipated in the island when applying a dc current $I_0$ reads $J_\mathrm{Q}=I_0^2/(4 t G_0)$. Furthermore, we relate the island's electron temperature increase $\DeltaTc=\Tc-T_0$ to an excess thermal noise $\Delta S$ measured at finite $I_0$: $\DeltaTc=\Delta S/(\kB t G_0)$~\cite{LeBreton2022,SM}. All measurements shown here were performed with the temperature of the dilution refrigerator fixed to $20~$mK. The $\DeltaTc(I_0)$ measurements are shown as symbols in Fig.~\ref{fig3}. Fig.~\ref{fig3}a shows the measurements for the \QH states $(2,0)$ and $(3,0)$ at $B=+10.5~$T. Fig.~\ref{fig3}b shows the measurements for the $(-2,-2$) \CI~and $(-3,-1.5)$ SBCI at $B=-10.5~$T. Fig.~\ref{fig3}c shows the measurements for the $(-4,-2)$ \CI~at $B=-7.2~$T, and Fig.~\ref{fig3}d shows the measurements for the \QH states $(1,0)$, $(2,0)$, $(3,0)$ and $(2,-4)$ at $B=+4~$T. All measurements show a V-shaped dependence, with a linear dependence at high $I_0$ whose slope decreases with $t$. The data match the predictions of the heat balance model of Eq.~\ref{eq:heatbal} (grey lines), with $J_\mathrm{Q}=I_0^2/(4 t G_0)$ and a corrective term due to heat Coulomb blockade (HCB) effects. In the HCB~\cite{Slobodeniuk2013,Sivre2018}, the large charging energy of the metallic island (estimated here to be $\EC/\kB\approx309~$mK, see Methods and Supplementary Information~\cite{SM}) prevents it from cooling by emitting net current fluctuations that would change its overall charge. This can suppress up to one ballistic channel if both $\Tc$ and $T_0$ are significantly smaller than $2N\EC/\kB$~\cite{Slobodeniuk2013,Sivre2018}, leading to a total quantized heat flow given by $(2N-1)\Jqe$ instead of $2N\Jqe$. The lines in Fig.~\ref{fig3} are obtained by calculating the HCB contribution with $N=\lvert t\rvert$, $\EC/\kB\approx309~$mK, and $T_0$ fixed to $20~$mK for all data except $(2,-4)$ for which $T_0$ is set to $36~$mK. The elevated $T_0$ for the latter is attributed to a small gate leakage current which heats up the sample even at zero $I_0$ (see Methods and Supplementary Information~\cite{SM}).

The quantization of heat flow can be checked by replotting the data of Fig.~\ref{fig3} to show $\Jq/(0.5\kappa_0)$ as a function of the difference $(\Tc^2-T_0^2)$, as is done in Fig.~\ref{fig4}. In this representation, quantized heat flows appear as straight lines with an integer slope given by the number of ballistic channels carrying heat away from the metallic island. Remarkably, all data corresponding to states with equal $t$, regardless of the value of $s$ (and thus on the \QH versus \CI~nature) fall on the grey full line given by $(2N-1)$, with $N=t$, signaling a quantized heat flow with fully developed HCB for all states. This is confirmed by the HCB predictions, shown as dark grey dashed lines, which are not distinguishable from the $(2N-1)$ lines for $N\geq2$. For $N=1$, $2N\EC/\kB T_0$ is not large enough for HCB to be fully developed even at the lowest temperatures~\cite{Slobodeniuk2013,Sivre2018,SM}, leading to a slightly increased heat flow that matches the data.

Our measurements demonstrate that the various topological states of the Hofstadter butterfly have a quantized heat flow set by their Chern number. Furthermore, we show that their transport properties, including the more subtle effects such as HCB, are virtually indistinguishable from that of \QH states, highlighting the paramount role of topology. Finally, our work  invites to investigate heat transport in the recently observed zero-field FCIs~\cite{Park2023,Lu2024,Lu2025}, which could host non-Abelian anyonic phases.

\section{Acknowledgments}

 This work was funded by the ERC (ERC-2018-STG QUAHQ), by the Investissements d’Avenir LabEx PALM (ANR-10-LABX-0039-PALM), by the ANR (ANR-24-CE47-2695 NONABEG and ANR-23-CE47-0002 CRAQUANT), and by the Region Ile de France through the DIM QUANTIP. K.W. and T.T. acknowledge support from the JSPS KAKENHI (Grant Numbers 21H05233 and 23H02052) , the CREST (JPMJCR24A5), JST and World Premier International Research Center Initiative (WPI), MEXT, Japan. FDP warmly thanks C. Gripe and the technical services at LPENS for their support.


\section{Competing Interests}
The authors declare no competing interest.

 
\bibliography{moireheat}

\section{Methods}

\subsection{Device fabrication}

The moiré heterostructure is realized by using a single 50~nm-thick hBN crystal with clear straight edges that has been broken in two parts during the exfoliation process. The largest part is used as the top hBN and is aligned with a clear edge of the graphene flake. The second part is used as the bottom hBN, and is aligned at $30~^\circ$ with the graphene flake, so as to ensure that the latter is crystallographically aligned with one of the two hBNs. A thin graphite flake is used as a back gate. Side metallic contacts are made by $\mathrm{CHF}_3 / \mathrm{0}_2$ reactive ion etching of the stack (70~nm deep, leaving 30 nm of the bottom hBN), followed by Cr/Pd/Au metal deposition in a few $10^{-8}~$mbar vacuum. The stack is then etched a second time to define the device geometry. We deliberately keep the overall size of the device small ($\leq5\times5$~\unit{um}, see Fig.~\ref{fig1}b) to avoid angle inhomogeneities that would give rise to a non-uniform quantized state across the sample.

 \subsection{Electrical conductance measurements}

 Measurements are performed in a cryogen-free dilution refrigerator with base temperature 9~mK, under high magnetic fields (up to 14~T) obtained with a superconducting magnet. The measurement lines are heavily filtered to obtain low base electron temperatures (see details of the wiring in the Supplementary Information~\cite{SM}). The electrical conductance is extracted from lock-in measurements at frequencies below 10~Hz. The transconductances between the various terminals of the device are measured simultaneously using different, non-commensurate frequencies. We measure the ac voltage drop on the various electrodes of the sample after amplification using CELIANS EPC1B room temperature amplifiers. In particular, whenever the sample is in a quantized state with the chirality indicated in Fig.~\ref{fig1}a, the equal current splitting leads to a voltage drop on the central island $dV_\mathrm{c}=I_0/(2tG_0)$. Current chirality then ensures that the measured ac voltages $dV_R$ and $dV_T$ are equal to $dV_\mathrm{c}$, leading to the transconductances $(dV_{R,T}/dI_0)^{-1}=(dV_\mathrm{c}/dI_0)^{-1}=2tG_0$. All conductances were checked to be independent on the dc current $I_0$ (see Supplementary Information~\cite{SM}).

\subsection{Noise and thermal transport measurements}

The thermal noise of the sample is detected through two independent measurement lines, labeled A and B, shown in Fig.~\ref{fig1}a. Each consists of a $RLC$ resonator placed in parallel to the sample, that filters the noise around $840~$kHz, followed by an amplification chain. The latter is made of a homemade low-noise voltage preamplifier anchored to the still plate of the dilution refrigerator (voltage gain approx. $5.5$) and a room temperature amplifier (NF SA-220F5, voltage gain $400$), after which both lines are sent to a digitizer. The $RLC$ resonator is realized by an approx. $190~$\unit{\uH} shunting inductor thermalized at the mixing chamber stage of the dilution refrigerator, with the capacitance $C\approx190~$pF being that of the coaxial lines connecting the sample to the low temperature amplifier, and the resistance $R\approx50~$k$\Omega$ modeling the losses in the circuit in absence of the sample. We record both autocorrelation signals A$\times$A and B$\times$B, as well as the cross correlation A$\times$B. The excess thermal noise $\Delta S$ is computed from all three signals to remove spurious noise contributions~\cite{LeBreton2022,SM}: $\Delta S=(($A$\times$A$+$B$\times$B$)/2-$A$\times$B$)/2$. The measurement lines were calibrated (overall gain and $RLC$ parameters) on a regular basis, typically weekly, by finely measuring the temperature dependence of the resonance spectra for all quantized states at a given magnetic field. The calibration procedure is described in details in the Supplementary Information~\cite{SM}. The overall gains of both lines were found to be constant and stable within $5~\%$ during the duration of the experiment (about 5 months), including after bias and thermal cycling above 100~K.

All measurements shown here are performed at a dilution refrigerator temperature regulated to $20~$mK. This allows us to fix the base electron temperature to $T_0=20~$mK for all datasets, except $(2,-4)$ at $B=4~$T where $T_0-38~$mK. This increase is confirmed when comparing the base value $S(I_0)$ of the noise at $I_0=0$ between $(2,-4)$ and $(2,0)$ while keeping the magnetic field and fridge temperature constant, by measuring the noise as a function of the gate voltage. We show in the Supplementary Information~\cite{SM} the noise versus $\Vg$ measurements for all considered topological states. All show clear plateaus in the noise, matching those in the conductance, except for $(2,-4)$ where the noise clearly increases with $\lvert \Vg\rvert$, such that $S_{(2,-4)}(I_0)-S_{(2,0)}(I_0)= 4\times2G_0 \kB\Delta T_0$, with $\Delta T_0=15\pm1~$mK. This confirms that $T_0$ is higher for $(2,-4)$, likely due to the finite gate leakage current flowing into the sample and generating additional heating (see leakage current measurements in the Supplementary Information~\cite{SM}).

The robustness of the results was checked by repeating the measurements at different fridge temperatures (up to $50~$mK), and various magnetic fields ($B=-11~$T, $6$~T, $4.2~$T). We show them in the Supplementary Information~\cite{SM}, demonstrating the same heat flow quantization as the data shown here, including the effects of HCB. In addition, we checked that our results are independent of $\Vg$ on all probed plateaus by measuring the noise at zero and finite $I_0$ as a function of $\Vg$ (see Supplementary Information~\cite{SM}).

\subsection{Heat Coulomb blockade}

In presence of $2N$ ballistic channels, HCB reduces the overall electronic heat flow from the central island with charging energy $\EC$ by up to one channel according to the formula~\cite{Slobodeniuk2013,Sivre2018}:
\begin{multline}
J_\mathrm{HCB}^\mathrm{e}=
2N\frac{\kappa_0}{2}(\Tc^2-T_0^2)\\
+\frac{(2N)^2\EC^2}{\pi^2h}\left[I\left( \frac{2N\EC}{\pi \kB T_0}  \right) -I\left( \frac{2N\EC}{\pi \kB\Tc} \right) \right], \label{eq:HCB}
\end{multline}
with
\begin{equation}
I(x)=\frac{1}{2}\left[ \ln\left(\frac{x}{2\pi}\right) - \frac{\pi}{x} -\psi\left( \frac{x}{2\pi} \right) \right],\label{eqI}
\end{equation}
where $\psi(z)$ is the digamma function. The grey lines in Fig.~\ref{fig3} and Fig.~\ref{fig4} are computed using Eq.~\ref{eq:HCB}, with the estimation of $\EC\approx309~$mK based on a plane capacitor model of the capacitance between the $1~$\unit{\um}$~\times~3~$\unit{\um} central metallic island and the graphite back gate, separated by $30~$nm of hBN. $\EC$ can also be extracted by fitting the data for $N=1$, where both the temperature dependence and the relative effect of HCB are the strongest. We performed measurements for the $(1,0)$ state at $B=+6~$T at fridge temperatures ranging from 20 to 50~mK (see Supplementary Information~\cite{SM}), yielding a value $\EC\approx319\pm87~$mK in good agreement with our geometrical estimate.

\clearpage


\section*{Supplementary material}

# Supplementary Information for "Quantized heat flow in the Hofstadter butterfly"

A. Zhang,[1] G. Aissani,[2] Q. Dong,[3] Y. Jin,[4] K. Watanabe,[5] T. Taniguchi,[6] C. Altimiras,[1] P. Roche,[1]
J.-M. Berroir,[2] E. Baudin,[2] G. Fève,[2] G. Ménard,[2] O. Maillet,[1] and F.D. Parmentier[1, 2, *]

[1]*Université Paris-Saclay, CEA, CNRS, SPEC, 91191 Gif-sur-Yvette cedex, France*
[2]*Laboratoire de Physique de l'Ecole normale supérieure, ENS, Université PSL,*
*CNRS, Sorbonne Université, Université Paris Cité, F-75005 Paris, France*
[3]*CryoHEMT, 91400 Orsay, France*
[4]*Université Paris-Saclay, CNRS, Centre de Nanosciences et de Nanotechnologies (C2N), 91120 Palaiseau, France*
[5]*Research Center for Electronic and Optical Materials,*
*National Institute for Materials Science, 1-1 Namiki, Tsukuba 305-0044, Japan*
[6]*Research Center for Materials Nanoarchitectonics,*
*National Institute for Materials Science, 1-1 Namiki, Tsukuba 305-0044, Japan*
(Dated: January 9, 2026)

## MEASUREMENT CIRCUIT

Supplementary Fig. 1 shows the schematics of the conductance measurement circuit. All lines are heavily filtered and thermalized at the mixing chamber stage of our dilution refrigerator. The filters are cascaded RC filters, with $R \approx 1$ k$\Omega$ and $C \approx 400$ nF. The current feed line (shown in red) is a dedicated line, with additional filtering (typically $C \approx 800$ nF) as well as a 1 M$\Omega$ bias resistor thermalized to the 3.5 K stage of the refrigerator. The $V_1$ measurement line, shown in dark green, has an approx. 800 $\Omega$ unintentional shunt to the ground at the level of the sample (likely due to a leak in the wiring). We have thus decided to use it solely to further characterize the device instead of using it to heat up the sample in a fashion symmetric to the $I_0$ current feed line.

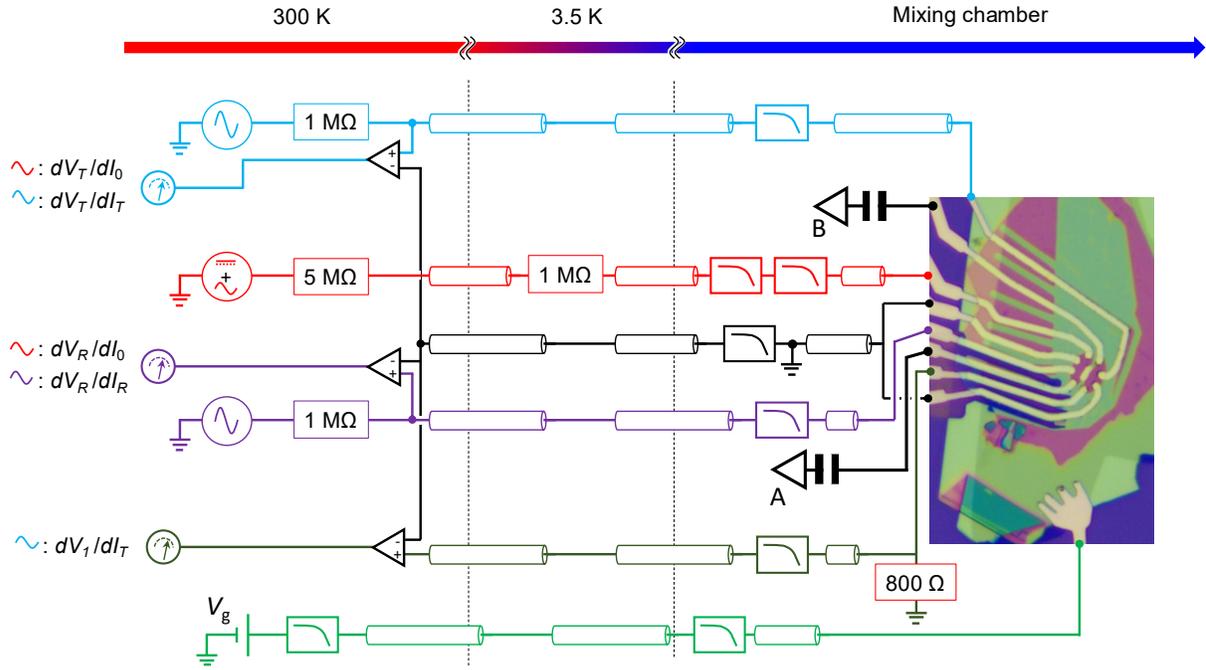

**Supplementary Figure 1** | Measurement circuit diagram.

## CONDUCTANCE MEASUREMENTS

### Raw conductance data and alignment angle extraction

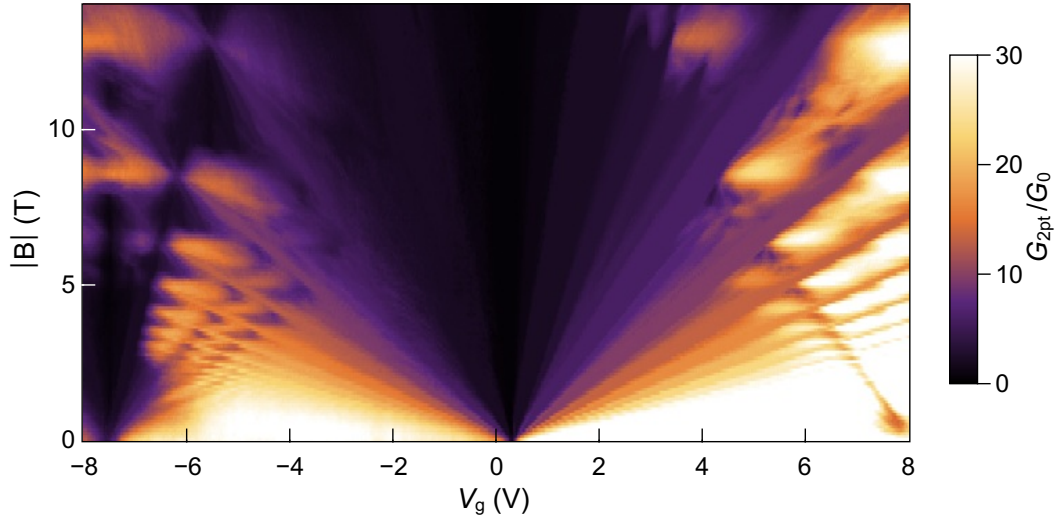

**Supplementary Figure 2** | 2-point differential conductance $(dV_T/dI_T)^{-1}$ measured versus $V_g$ and $|B|$, at $T \approx 20$ mK.

The raw measurement of the 2-point conductance $(dV_T/dI_T)^{-1}$ versus gate voltage and magnetic field is shown in Supplementary Fig. 2. From this measurement, one can extract through two methods the moiré superlattice length $\lambda$ as well as the alignment angle $\theta$, the two being related through $\lambda = \frac{a(1+\delta)}{\sqrt{\delta^2 + 2(1+\delta)(1-\cos\theta)}}$ ($a = 0.246$ nm is the lattice constant of graphene, and $\delta = 0.017$ the lattice mismatch with hBN) [1–3]. First, the $V_g$ position of the charge neutrality point (CNP) and the satellite peaks allow computing the density moiré unit cell $n_0 = 2/(\sqrt{3}\lambda^2)$ through $n_{sat} = C_g/e \times \Delta V_{sat} = 4n_0$, where $n_{sat}$ is the density position of the satellite peak, $\Delta V_{sat} = V_{sat} - V_{CNP}$ the gate voltage distance between CNP and satellite peak, and $C_g$ is the gate capacitance per unit area.

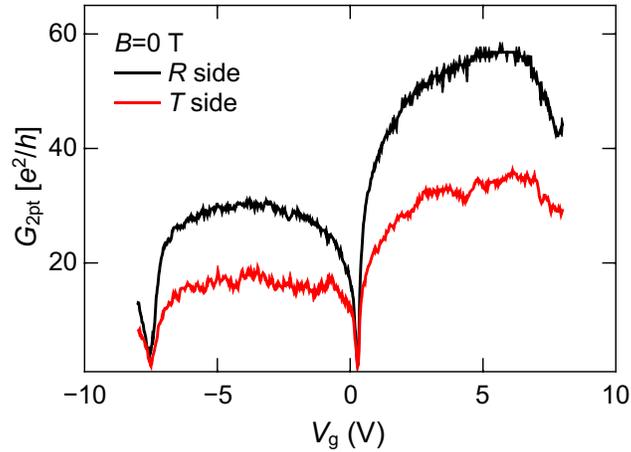

**Supplementary Figure 3** | 2-point differential conductance $(dV_R/dI_R)^{-1}$ (black) and $(dV_T/dI_T)^{-1}$ (red) versus $V_g$, at $B = 0$ T and $T \approx 20$ mK.

Supplementary Fig. 3 shows the $B = 0$ T measurement of the 2-point conductances on both sides of the sample versus $V_g$, allowing to extract $\Delta V_{sat} = 7.86$ V. With a hBN thickness of 50 nm, his leads to $\lambda = 12.54$ nm and $\theta = 0.59$ °.

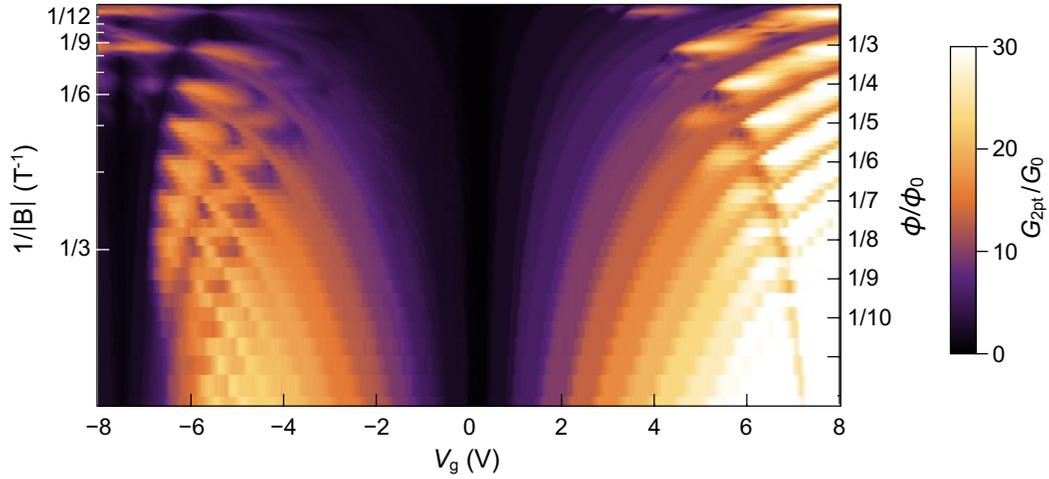

**Supplementary Figure 4** | 2-point differential conductance $(dV_T/dI_T)^{-1}$ plotted versus $V_g$ and $1/|B|$.

The second method consists in tracking the periodicity of the CNP clones by replotting the conductance maps as a function of $1/|B|$. This is shown in Supplementary Fig. 4, showing clear periodic features with frequency $B_1 \approx 25.8$ T. This value corresponds to $\phi/\phi_0 = (B \times \sqrt{3}\lambda^2/2)/(h/e) = 1$, yielding $\lambda = 13.6$ nm and $\theta = 0.4$ °. The lines shown in the $n/n_0$, $B$ plots of main text Fig. 1 as well as the following plots are calculated using this second value.

**Additional maps**

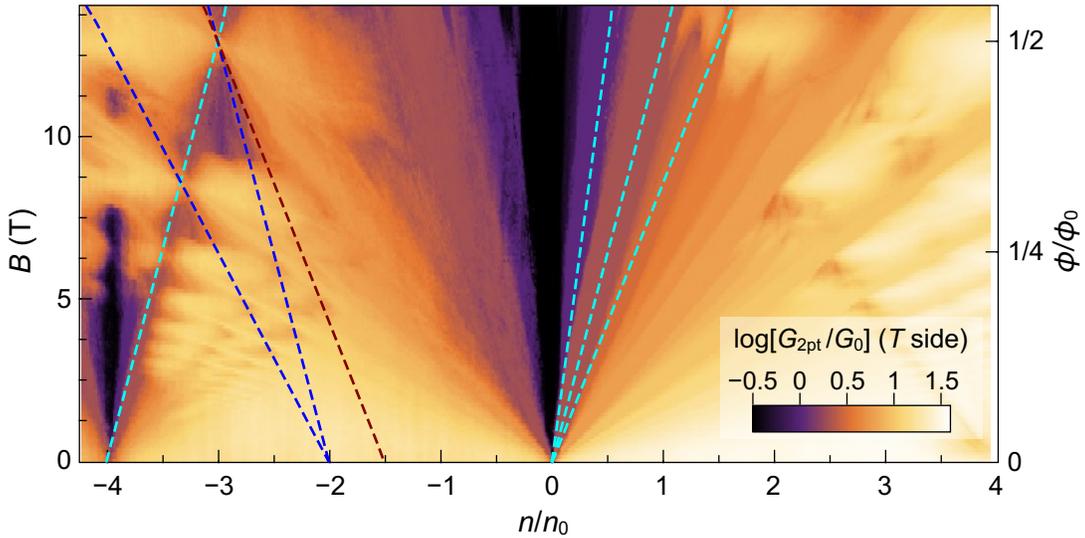

**Supplementary Figure 5** | $(dV_T/dI_T)^{-1}$ plotted versus $n/n_0$ and $|B|$. The dashed lines correspond to those shown in the main text.

The log-scale plot of the 2-point conductance on the transmitted side $(dV_T/dI_T)^{-1}$, taken from the data shown in Supplementary Fig. 2, is shown in Supplementary Fig. 5 versus $n/n_0$ and $B$. The diophantine equation lines shown in main text Fig. 1 for $(dV_R/dI_R)^{-1}$ match well the features of $(dV_T/dI_T)^{-1}$, confirming the homogeneity of the sample.

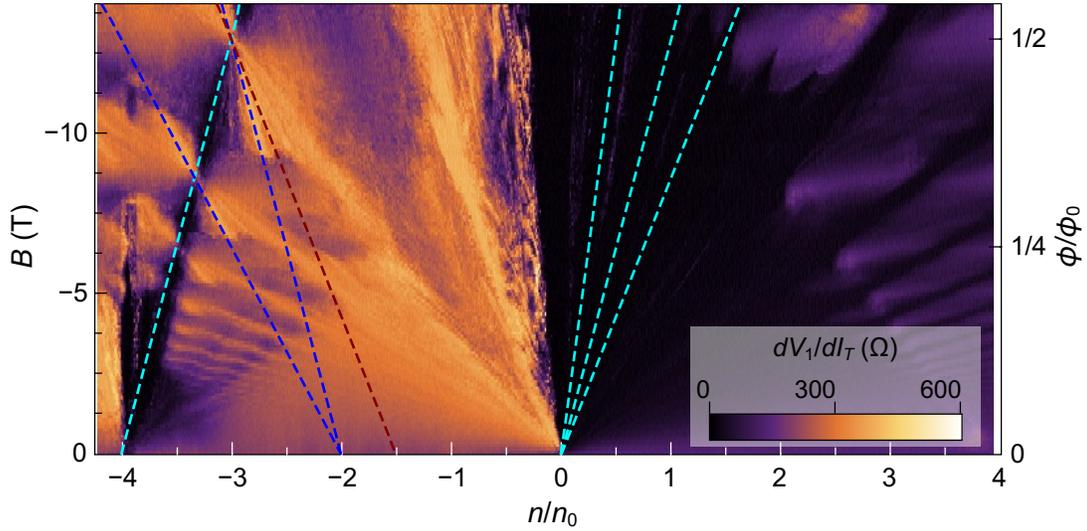

**Supplementary Figure 6** | $dV_1/dI_T$ plotted versus $n/n_0$ and $B$. At negative $B$, the chirality is such that for positive $t$, the current $dI_T$ directly flows into the cold ground downstream of the contact $T$, leading to vanishing $dV_1$. On the other hand, negative $t$ lead to a finite $dV_1$ (see e.g. along the blue dashed lines corresponding to $(-4, -2)$ and $(-2, -2)$).

The chirality of charge transport is further checked by measuring the various transconductances for a sign of the magnetic field such that the chirality is opposite to that shown in main text Fig. 1. In that case the signal vanishes since all current is directly sent to the nearest cold ground. This illustrated in Supplementary Fig. 6 for $dV_1/dI_T$ at negative $B$, and in Supplementary Fig. 7 for $dV_T/dI_0$ and $dV_R/dI_0$ near the $\phi/\phi_0 = 1/3$ and $\phi/\phi_0 = 1/2$ CNP clones at positive $B$.

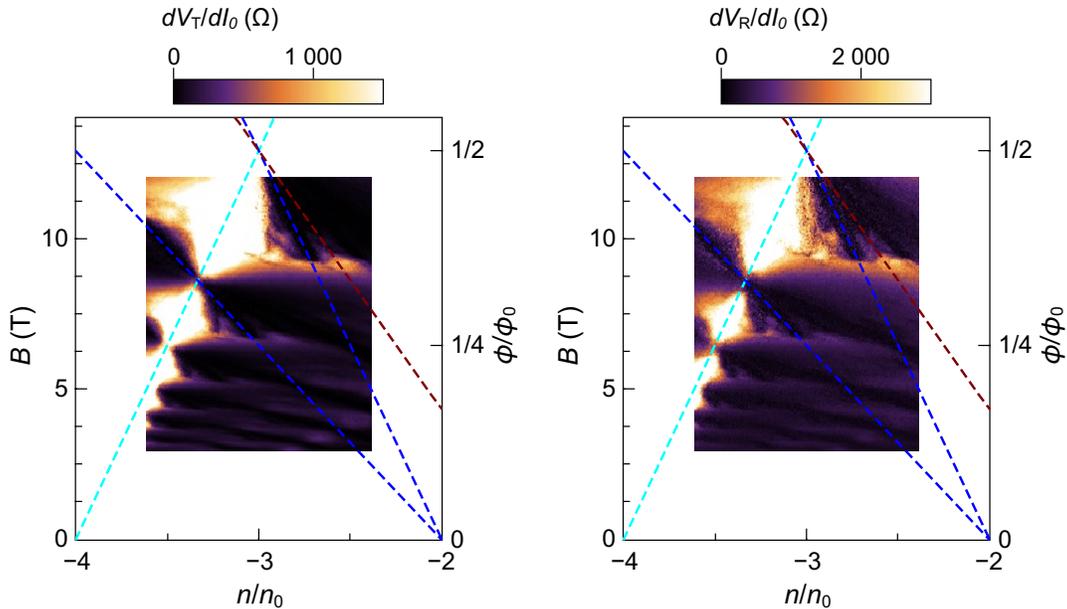

**Supplementary Figure 7** | $dV_T/dI_0$ (left) and $dV_R/dI_0$ (right) versus $n/n_0$ and $B$. For positive $B$, the chirality is such that for negative $t$, the current $dI_0$ directly flows into the cold ground downstream of the current feed contact, leading to vanishing $dV_T$ and $dV_R$.

The conductance versus $n/n_0$ linecut at $B = +10.5$ T, completing the data shown in main text Fig. 2, is shown in Supplementary Fig. 8. Note that while the 2-point conductances are equal and well quantized for the $(1,0)$ state

at $n/n_0 \approx 0.4$, the transmitted and reflected transconductances are unequal and far from quantized, demonstrating improper current splitting at the central island for this state at higher magnetic field.

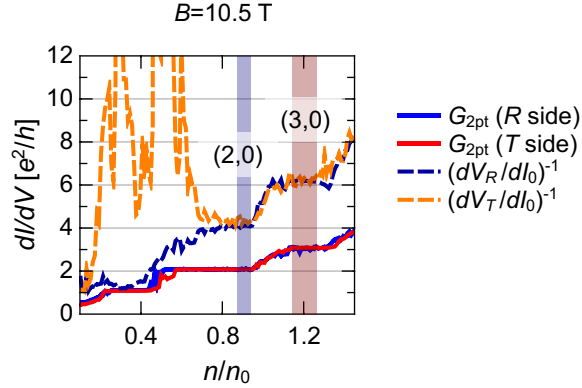

**Supplementary Figure 8** | Measured differential conductances versus $n/n_0$ at $B = +10.5$ T for $T = 20$ mK. Full lines: 2-point conductances $(dV_R/dI_R)^{-1}$ (blue) and $(dV_T/dI_T)^{-1}$ (red). Dashed lines: reflected (dark blue) and transmitted (orange) transconductances $(dV_{R,T}/dI_0)^{-1}$.

### Conductance versus $I_0$

An important prerequisite of the thermal transport measurement is that the electrical conductance is independent of the dc current $I_0$. We show in Supplementary Fig. 9 the measurements of the 2-point conductances as well as of the transmitted and reflected transconductances versus $I_0$, for all topological states probed in the main text. All show constant values (up to some enhanced fluctuations in $(dV_R/dI_0)^{-1}$ for some states, notably $(-3, -1.5)$; these fluctuations are not reproducible), indicating good current splitting even at finite current bias.

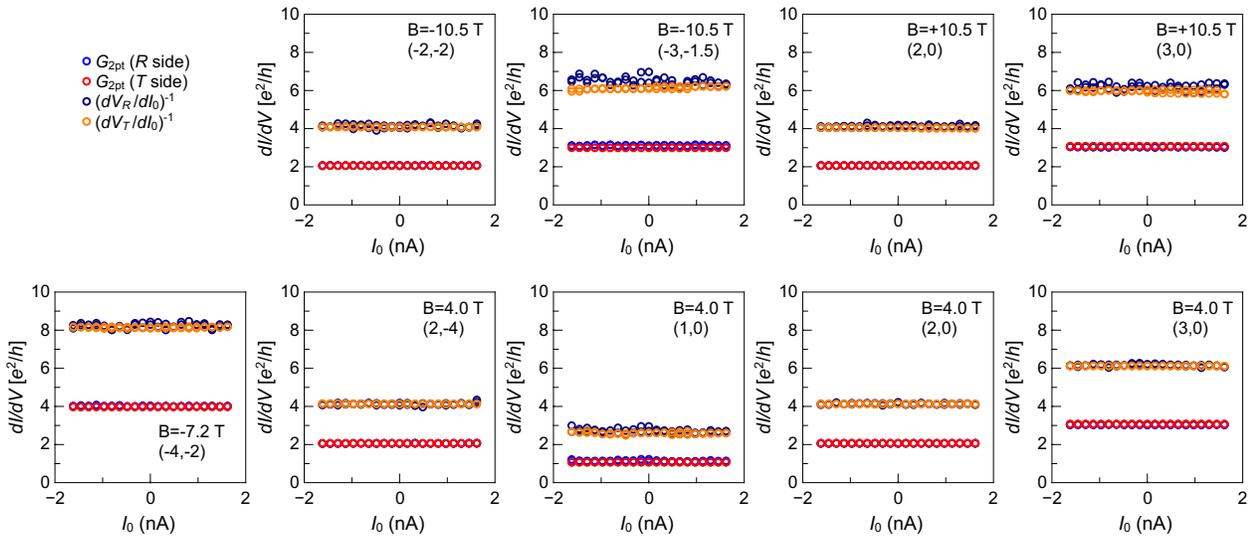

**Supplementary Figure 9** | Differential conductances $(dV_R/dI_R)^{-1}$ (blue), $(dV_T/dI_T)^{-1}$ (red), $(dV_R/dI_0)^{-1}$ (dark blue) and $(dV_T/dI_0)^{-1}$ (orange) versus $I_0$, for all topological states probed in the main text.

## NOISE MEASUREMENTS

### Setup and calibration

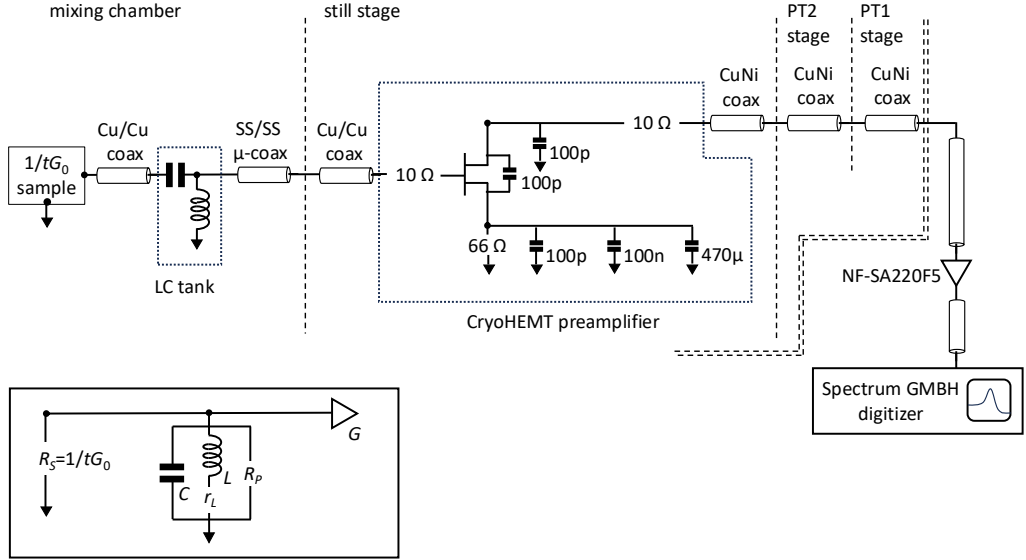

**Supplementary Figure 10** | Schematics of the noise measurement setup. Inset: equivalent circuit used in the calibration.

The noise measurement setup is based on two identical measurements lines depicted in Supplementary Fig. 10. They include a RLC resonator materialized by a 190 µH copper wound shunt inductor thermalized to the mixing chamber stage, and a series of amplification stages: a custom made low-temperature voltage preamplifier (gain approx. 5) based on a CryoHEMT low-noise transistor, and a room-temperature NF-SA22F5 follower amplifier (gain 400). The equivalent circuit is shown in the inset of Supplementary Fig. 10: the capacitor $C$ represents the distributed capacitance of the coaxial lines connecting the sample to the low-temperature preamplifier, the parallel resistor $R_P$ corresponds to the losses in the circuit (partly made of resistive material to avoid thermal shunts between the different stages of the refrigerator), $r_L \approx 20~\Omega$ corresponds to the resistance of the inductor, and $G$ is the compound voltage gain of the two cascaded amplifiers. The 20 nF coupling capacitor is not included in the equivalent circuit, as it is essentially a shunt at the approx. 0.9 MHz RLC resonance frequency.

After amplification, the voltage in each line recorded by a dual input Spectrum GMBH digitizer board with which we compute the auto and cross spectra $S_v^{AA}(f)$, $S_v^{BB}(f)$, $S_v^{AB}(f)$ that are then averaged over a frequency bandwidth $\Delta f = [865 - 900]$ kHz encompassing the RLC tank resonance. To extract the current noise of the sample $S_{i,sample}$ from the averaged auto and cross spectra $\bar{S}_v^{AA}$, $\bar{S}_v^{BB}$, $\bar{S}_v^{AB}$, one need to accurately describe and determine all the parameters of the amplification chain. We start from the following expressions:

$$\bar{S}_v^{AA} = \frac{1}{\Delta f}\int_{\Delta f} S_v^{AA}(f) = G_A^2 \times \frac{1}{\Delta f}\int_{\Delta f} df\left[S_{v,amp}^A + \left|Z_{//}^A\right|^2\left(S_{i,amp}^A + 4k_B T Re\left(\frac{1}{Z_{RLC}^A}\right) + S_{i,sample}\right)\right] \quad (1)$$

$$\bar{S}_v^{BB} = \frac{1}{\Delta f}\int_{\Delta f} S_v^{BB}(f) = G_B^2 \times \frac{1}{\Delta f}\int_{\Delta f} df\left[S_{v,amp}^B + \left|Z_{//}^B\right|^2\left(S_{i,amp}^B + 4k_B T Re\left(\frac{1}{Z_{RLC}^B}\right) + S_{i,sample}\right)\right] \quad (2)$$

$$\bar{S}_v^{AB} = \frac{1}{\Delta f}\int_{\Delta f} S_v^{AB}(f) = -G_A G_B \times \frac{1}{\Delta f}\int_{\Delta f} df\left[Z_{//}^A\left(Z_{//}^B\right)^* S_{i,sample}\right] \quad (3)$$

with $S_{i,amp}^{A/B}$ and $S_{v,amp}^{A/B}$ the current and voltage noises of the amplifier A or B, $4k_B T Re(\frac{1}{Z_{RLC}^{A/B}})$ the thermal noise of the RLC resonator, $Z_{//}^{A/B}$ the parallel impedance of the RLC circuit resonator and the sample, $\Delta f$ the integration

bandwidth, and $S_{i,sample}$ the current noise of the sample, which includes its thermal noise written as $4k_BT \times t \times e^2/h$. To determine the parameters of the RLC circuit resonator as well as the voltage and current noise of the preamplifiers, we use a temperature calibration method. We measure the equilibrium noise between 40 to 200 mK at different $t$. Supplementary Fig. 11 and Supplementary Fig. 12 show typical raw spectra measured for the calibration.

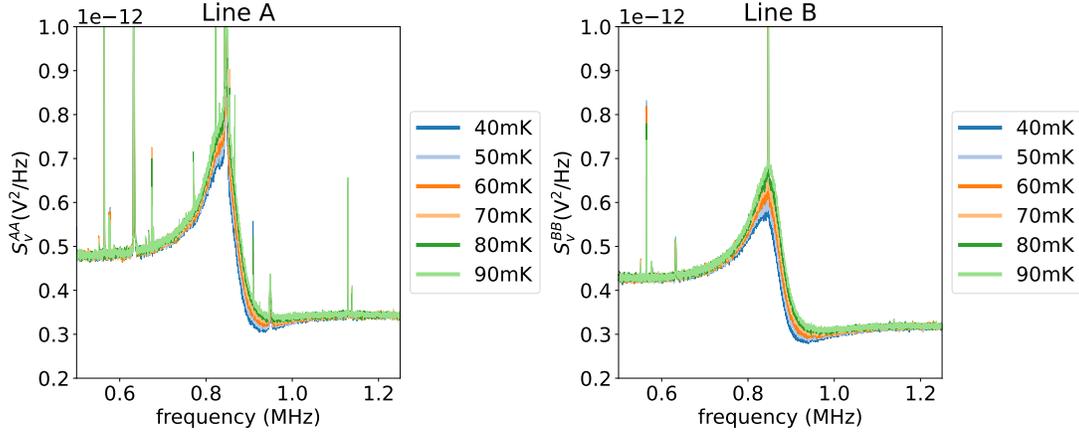

**Supplementary Figure 11** | Raw noise spectra measured for fridge temperature ranging from 40 mK to 90 mK for the $(2,0)$ state at $B = 4$ T. Left: A×A; right: B×B.

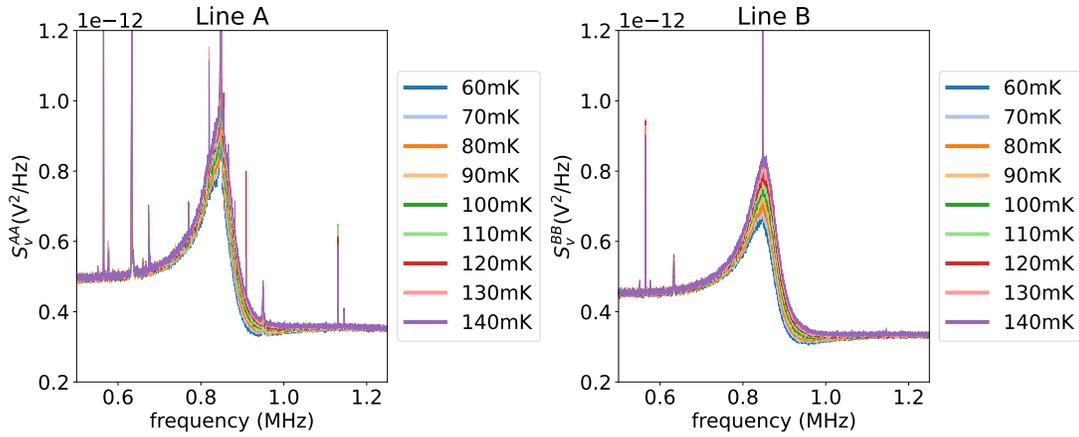

**Supplementary Figure 12** | Raw noise spectra measured for fridge temperature ranging from 60 mK to 140 mK for the $(2,0)$ state at $B = +10.5$ T. Left: A×A; right: B×B.

According to Supplementary Eq. 3, the noise includes a temperature-independent part due to the amplifier's noise. In order to first extract the values of the gain and the RLC circuit parameters, we remove to each spectrum a reference spectrum made of the average of all spectra for the different measured temperatures. We restrict the temperature range to typically $[50 - 120]$ mK, where the integrated noise versus temperature (see Supplementary Fig. 15 and Supplementary Fig. 16) remains linear.

$$\Delta S_v^{AA/BB}(f,T) = S_v^{AA/BB}(f,T) - \frac{1}{N_T}\sum_{k=1}^{N_T} S_v^{AA/BB}(f,T_k) \tag{4}$$

The obtained spectra, examples of which are shown in Supplementary Fig. 13 and Supplementary Fig. 14, are then fitted by the expression below:

$$\Delta S_v^{AA/BB}(f) = G_{A/B}^2 \ 4k_B\Delta T \left|Z_{//}^{A/B}(f)\right|^2 \left[Re\left(\frac{1}{Z_{RLC}^{A/B}(f)}\right) + t \times e^2/h\right] \tag{5}$$

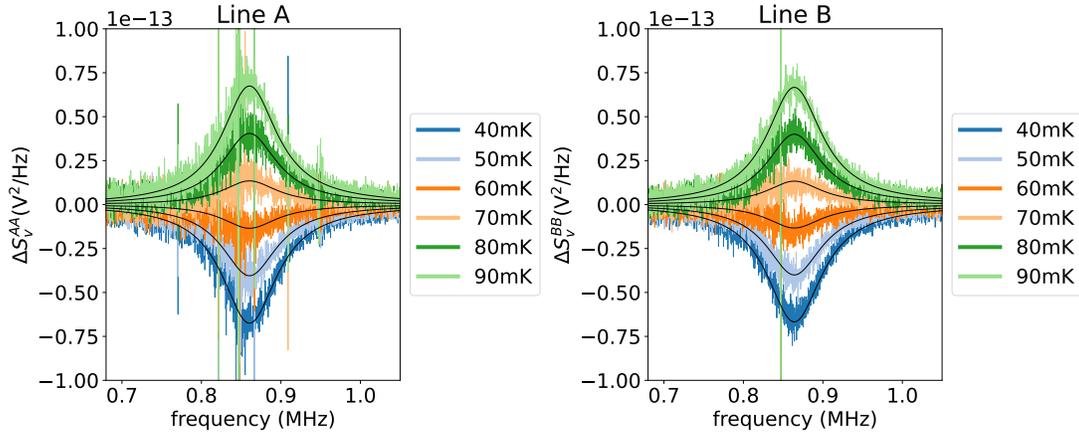

**Supplementary Figure 13** | Differential noise spectra, for fridge temperature ranging from 40 mK to 90 mK for the $(2,0)$ state at $B = 4$ T. Left: A×A; right: B×B. Black lines: fits using Supplementary Eq. 4.

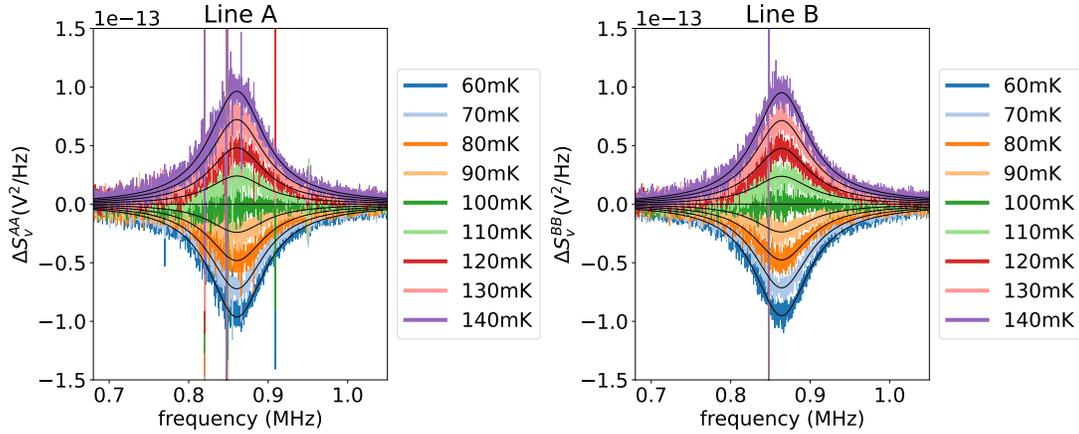

**Supplementary Figure 14** | Raw noise spectra measured for fridge temperature ranging from 60 mK to 140 mK for the $(2,0)$ state at $B = +10.5$ T. Left: A×A; right: B×B.

Fitting the obtained spectra using Supplementary Eq. 5 yields the parameter $R_{A/B} \approx 500/500$ kΩ, $C_{A/B} \approx 188.0/188.5$ pF, $L_{A/B} \approx 182/180$ uH, $r_{L_{A/B}} \approx 22/20$ Ω described in Supplementary Fig. 10 and the gain $G_{A/B} \approx 2340/2330$. The resulting fits are illustrated in Supplementary Fig. 13 and Supplementary Fig. 14. This calibration can then be applied to the measurements, by simply dividing the integrated noise measured at any given gate or bias voltage by a fixed conversion factor which only depends on the topological invariant $t$:

$$S_{i,sample} = \frac{\bar{S}_v^{AA/BB}}{\frac{G_{A/B}^2}{\Delta f} \int_{\Delta f} df \left| Z_{//}^{A/B}(f,t) \right|^2} \tag{6}$$

$$= \frac{\bar{S}_v^{AB}}{-\frac{G_A G_B}{\Delta f} \int_{\Delta f} df \left[ Z_{//}^A(f,t) \left( Z_{//}^B(f,t) \right)^* \right]} \tag{7}$$

The total voltage noise of the amplifier, including the part originating from back flowing current noise through the RLC circuit, can be found from the intercept in temperature dependence of the integrated noise before amplifier $\bar{S}_{v,norm}^{AA/BB}/G_{A/B}^2$, as illustrated in Supplementary Fig. 15 and Supplementary Fig. 16. Note that this plot makes apparent the fact that the resonator and sample temperature start differing from the fridge temperature below 50 mK. This is due to the fact that the thermalization times become significantly larger than the 30 minutes wait time used in the calibration procedure (note that this is even more pronounced for the sample temperature $T_0$: to obtain $T_0 = T = 20$ mK, we typically wait at least 72-96 hours, depending on the magnetic field, before performing the

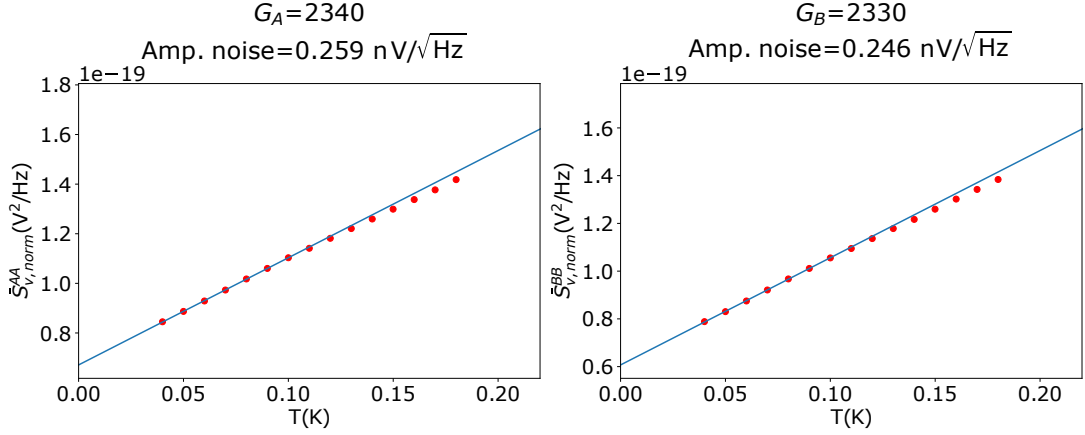

**Supplementary Figure 15 |** Integrated noise versus fridge temperature for the $(2,0)$ state at $B = 4$ T. Left: A×A; right: B×B. Lines are linear fits, yielding the amplifier noise at $T = 0$.

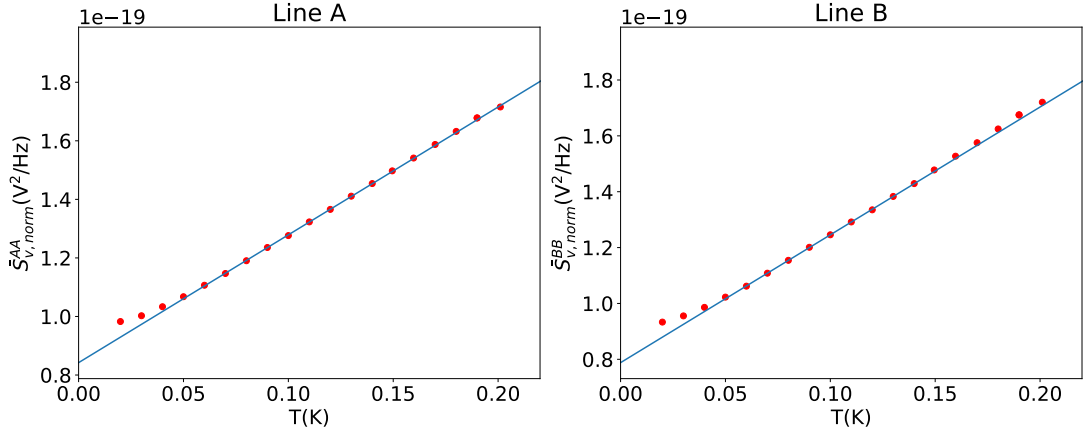

**Supplementary Figure 16 |** Integrated noise versus fridge temperature for the $(2,0)$ state at $B = +10.5$ T. Left: A×A; right: B×B. Lines are linear fits, yielding the amplifier noise at $T = 0$.

noise versus $I_0$ measurements). At T = 0 K, we omit the thermal noise from sample and RLC circuit resonator, leaving only the contribution of the amplifier, estimated to $S_{tot,amp}^{A/B} \approx 0.26/0.25$ nV/$\sqrt{\text{Hz}}$. Note that this includes a finite current noise, as indicated by the asymmetry in the raw resonance spectra shown in Supplementary Fig. 11 and Supplementary Fig. 12, which corresponds to the crossover between voltage and current noise at the resonance, roughly proportional to the imaginary part of the RLC impedance [4].

### Auto and cross-correlations

We rely on our two noise measurement lines (A and B) to extract the thermal noise increase of the central metallic island independently of any additional noise source, *e.g.* shot noise generated at the current feed contact. The details of this method are explained in ref. [5]; in a nutshell, a finite electron temperature increase $\Delta T_c$ of the island leads to positive autocorrelations of the current fluctuations $\Delta S_{AA} = \Delta S_{BB} = t \times G_0 k_B \Delta T_c$, and negative (but with equal magnitude) cross-correlations $\Delta S_{AB} = -\Delta S_{AA,BB} = -t \times G_0 k_B \Delta T_c$, which corresponds to current conservation at the central island. Oppositely, a current noise incoming onto the central island will be evenly split, leading to a positive contribution in the cross-correlations, equal to that of the autocorrelation. Thus, calculating the weighted sum $((A×A+B×B)/2−A×B)/2$ allows suppressing any eventual incoming noise, while at the same time increasing the signal to noise ratio for the thermal noise. This is illustrated in Supplementary Fig. 17, showing the three signals as well as their weighted sum versus $I_0$ for the $(-2,2)$ state at $B = -10.5$ T. The cross-correlation has a slightly lower amplitude than the autocorrelations (and consequently, so has the weighted sum), indicating the presence of a

small input noise. Importantly, an additional shot noise generated at either the central island or at the measurement contacts would lead to different magnitudes between the two auto-correlation. Supplementary Fig. 17 illustrates that it is not the case in our experiment.

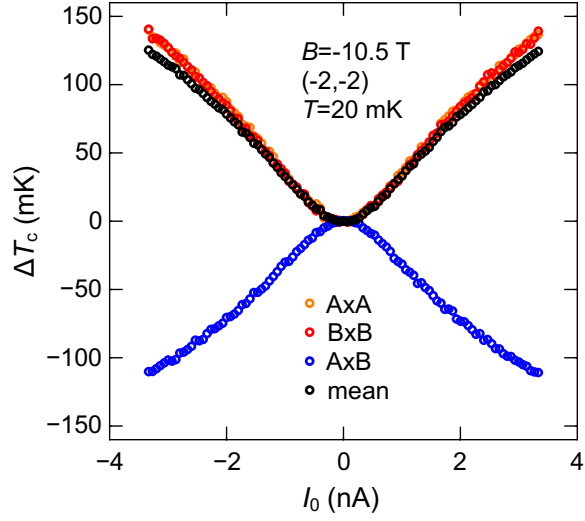

**Supplementary Figure 17** | Autocorrelation A×A (orange) and B×B (red), crosscorrelation A×B (blue) and mean signal ((A×A+B×B)/2−A×B)/2 (black) versus $I_0$, at $B = −10.5$ T and $T = 20$ mK for the $(−2, 2)$ state.

**Noise versus $V_g$ measurements**

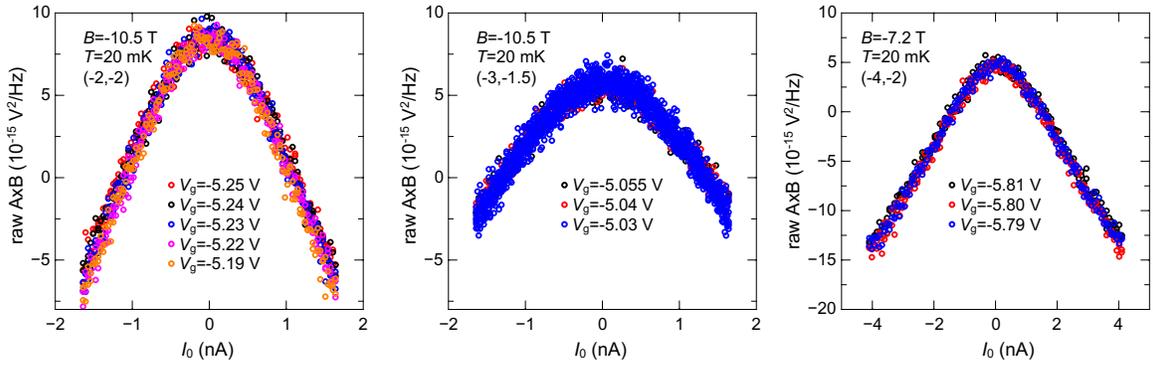

**Supplementary Figure 18** | Raw cross-correlation noise A×B versus $I_0$ for various $V_g$ for the $(−2, −2)$ state at $B = −10.5$ T (left), the $(−3, −1.5)$ state at $B = −10.5$ T (middle), and the $(−4, −2)$ state at $B = −7.2$ T (right). Measurements performed at $T = 20$ mK.

An important check in the experiment is to make sure that the heat transport results are consistent on a finite range of $V_g$ at a given magnetic field. This is shown in Supplementary Fig. 18, where we plot the raw cross-correlation noise measured as a function of $I_0$ for different values of $V_g$ for the states $(−2, −2)$, $(−3, −1.5)$ and $(−4, −2)$. The raw data fall on top of one another for excursions in $V_g$ exceeding 20 mV, indicating a plateau over which our results are constant. An additional check of the independence of the heat transport results reported in the main text can be done by measuring the noise versus gate finite at both zero and finite $I_0$. Supplementary Fig. 19 shows the measurement of the noise as a function of $V_g$ along the plateaus corresponding to all topological states investigated in the main text. The noise at both zero and finite $I_0$ show plateaus as a function of $V_g$ around the positions at which the results shown

in the main text were obtained, indicating that the latter are $V_g$-independent. In addition, the measurements shown in Supplementary Fig. 19 allow explaining the increase $T_0$ observed for the $(2, -4)$ state at $B = 4$ T: while the noise shows flat plateaus for all magnetic fields and all $|V_g| < 6$ V, it clearly increases with $|V_g|$ when the latter is pushed beyond 6 V in the $B = 4$ T data shown in Supplementary Fig. 19d. This increase of noise can be interpreted as an increased $T_0$, which can be extracted by comparing the noises at $V_g = -6.96$ V for $(2, -4)$ and at $V_g = 0.88$ V for $(2, 0)$ (vertical dashed lines in Supplementary Fig. 19). Doing so yields an increase $\Delta T_0 = 15 \pm 1$ mK, compatible with the 18 mK $T_0$ increased observed in the heat transport data.

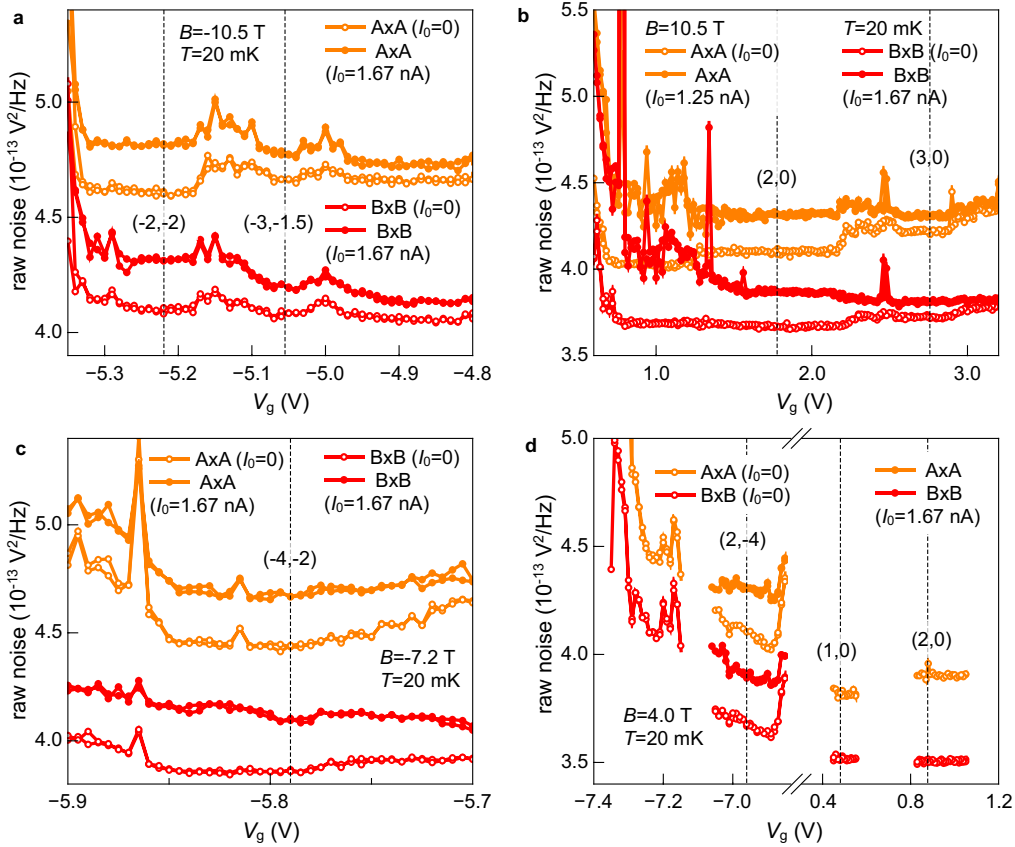

**Supplementary Figure 19** | Raw autocorrelation noise A×A (orange) and B×B (red) versus $V_g$, at zero $I_0$ (open symbols) and finite $I_0$ (full symbols). Measurements performed at $T = 20$ mK, for $B = -10.5$ T (**a**), $B = +10.5$ T (**b**), $B = -7.2$ T (**c**), and $B = +4$ T (**d**). The vertical dashed lines indicates where the thermal transport measurements shown in the main text were performed.

Supplementary Fig. 20 shows measurements of the gate leakage current, in presence and absence of the sample. While a significant portion of the leakage current is due to the refrigerator wiring, adding the sample yields an approx 5 nA increase in the leakage current, a fraction of which is likely to flow into the sample and lead to the increase in $T_0$. Note that a similar behavior was observed in a previous sample, studied in ref. [5].

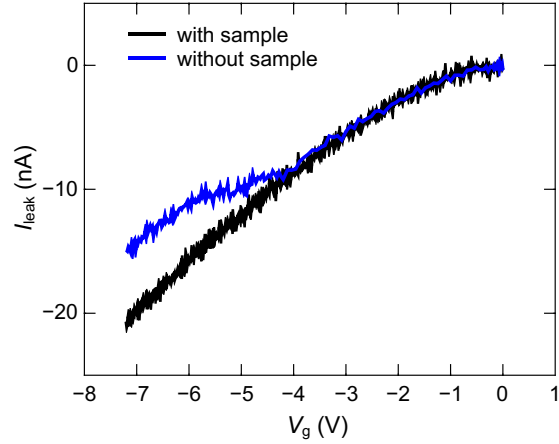

**Supplementary Figure 20 |** Leakage current measured versus $V_g$, at $B = 0$ $T$, in presence of the sample (black) at 30 mK, and without sample (blue) at 12 K.

### Temperature dependence

We have repeated the heat transport measurements at several fridges temperatures, leading to different values of $T_0$. Supplementary Fig. 21 shows the measurements for the $(-2, -2)$, $(-3, -1.5)$, and $(-4, -2)$ states investigated in the main text at $T_0 = 20$ mK (blue symbols) and $T_0 = 24$ mK (Supplementary Fig. 21a and b, red symbols) / $T_0 = 30$ mK (Supplementary Fig. 21c, red symbols). The data at higher $T_0$ is very well matched by the results of the simplified heat balance model with $(2N - 1)$ ballistic heat channels, indicating not only that our results are reproducible at higher $T_0$, but also that HCB is still fully developed at these temperatures. Importantly, increasing $T_0$ essentially affects the thermal rounding of the data at low bias, as the slope at large bias is only dictated by the topological invariant $t$, fixing the number of ballistic heat channels.

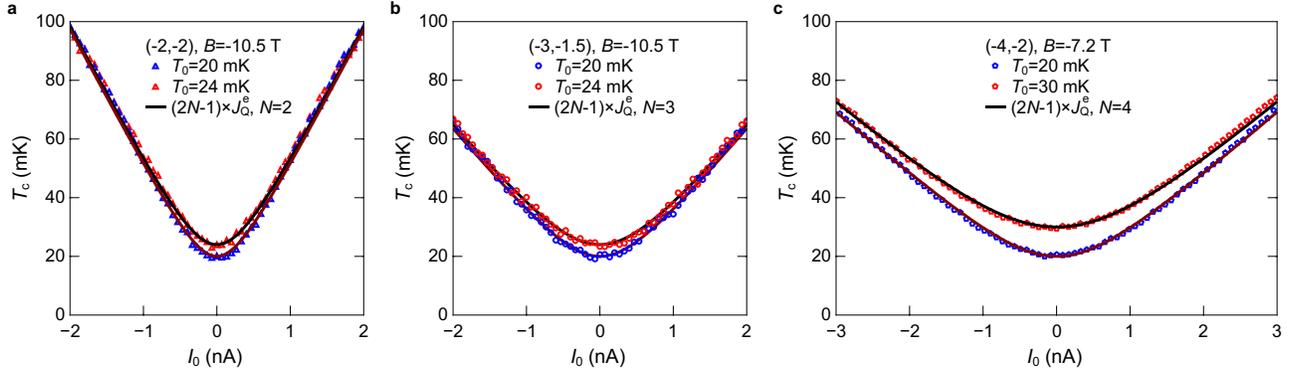

**Supplementary Figure 21 |** Measured $T_c$ versus $I_0$ for the $(-2, -2)$ state at $B = -10.5$ T (**a**), the $(-3, -1.5)$ state at $B = -10.5$ T (**b**), and the $(-4, -2)$ state at $B = -7.2$ T (**c**). Blue symbols: $T_0 = 20$ mK; red symbols: $T_0 = 24$ mK (**a** and **b**), $T_0 = 30$ mK (**c**). Black lines: prediction of the heat balance model without HCB, with $2N - 1$ ballistic heat channels.

The independence of the heat flow quantization with respect to $T_0$ is emphasized when plotting the data of Supplementary Fig. 21 in a $J_Q$ vs $T^2$ graph similar to that of main text Fig. 4. This is shown in Supplementary Fig. 22, demonstrating a quantized heat flow for all values of $T_0$, only set by $N = t$.

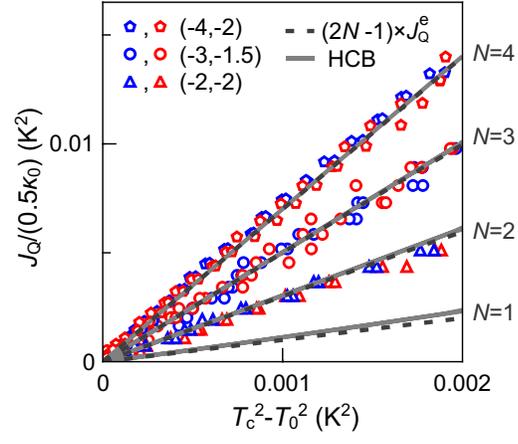

**Supplementary Figure 22** | Heat flow $J_Q/(0.5\kappa_0)$ versus $T_c^2 - T_0^2$. Symbols and colors correspond to the data shown in Supplementary Fig. 21 (blue: $T_0 = 20$ mK, red: $T_0 = 24/30$ mK). Dashed lines: quantized heat flow with full suppression of a single heat channel $(2N-1)J_Q^e$, with $N = 1 \rightarrow 4$ (from bottom to top) the number of ballistic heat channels on each side of the metallic island. Full lines: HCB predictions for the corresponding $N$ at $T_0 = 20$ mK.

### Heat Coulomb blockade

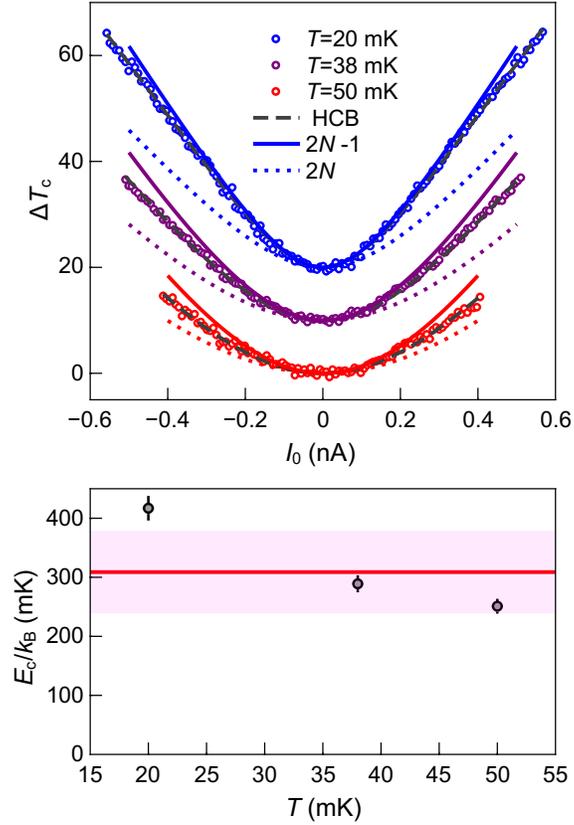

**Supplementary Figure 23** | Top: $\Delta T_c$ vs $I_0$ measured at $B = +6$ T for the $(1,0)$ state at $T = 20$ mK (blue), $T = 38$ mK (purple), and $T = 50$ mK (red). The 20 mK and 38 mK curves are shifted horizontally by (resp.) 20 and 10 mK for clarity. Symbols: experimental data. Dashed lines: heat balance including HCB, with $E_C$ as fit parameter. Full lines: heat balance without HCB, with $2N-1$ ballistic heat channels. Dotted lines: heat balance without HCB, with $2N$ ballistic heat channels. Bottom: symbols: values of $E_C$ extracted from the HCB fits of the top panel, plotted versus fridge temperature $T$. The red line corresponds to $E_C = 309$ $mK$, and the pink band to $\pm 70$ mK.

We show in Supplementary Fig. 23 heat transport data at $B = +6$ T for the $(1,0)$ state with which we provide an estimate of the island's charging energy $E_C$, independent of geometrical considerations. Because HCB is sensitive to the number of ballistic heat channels $2N$ [6,7], at low $N$, it is less likely to be fully developed, allowing to extract $E_C$ by fitting the $\Delta T_c$ vs $I_0$ data. We have performed the measurements for three values of the fridge base temperature $T = 20$ mK, 38 mK and 50 mK. At low temperature, the data is very close to a fully developed HCB, almost falling on the prediction of simple heat balance with $2N-1$ ballistic channels (full line). As the temperature increases, the data falls between the $2N-1$ prediction and the $2N$ one (dotted line), becoming closer to the latter at higher temperature. We have fitted the three datasets with the HCB prediction (dashed lines) with fixed $N = t = 1$ and $T_0 = T$, and $E_C$ a free parameter. The extracted values yield $E_C \approx 319 \pm 87$ mK, in good agreement with the geometrical estimate.

### ADDITIONAL HEAT TRANSPORT DATA

Supplementary Fig. 24 and Supplementary Fig. 25 show additional measurements performed for the states $(2,0)$ and $(3,0)$ at $B = +4.2$ T (Supplementary Fig. 24) and the states $(-2,-2)$ and $(-4,-1)$ at $B = -11$ $T$ (Supplementary Fig. 25), for various $T_0$. Here again, the agreement with the heat balance without HCB but with $2N-1$ ballistiuc heat channels is excellent, demonstrating quantized heat flow with a fully developed HCB. Note that the agreement is less good for

$(-4, -1)$ which shows a small but statistically significant increase in $T_c$ with respect to the predictions at $|I_0| > 1$ nA, which is likely due to a sub-optimal current splitting at the central island (see *e.g.* main text Fig. 2).

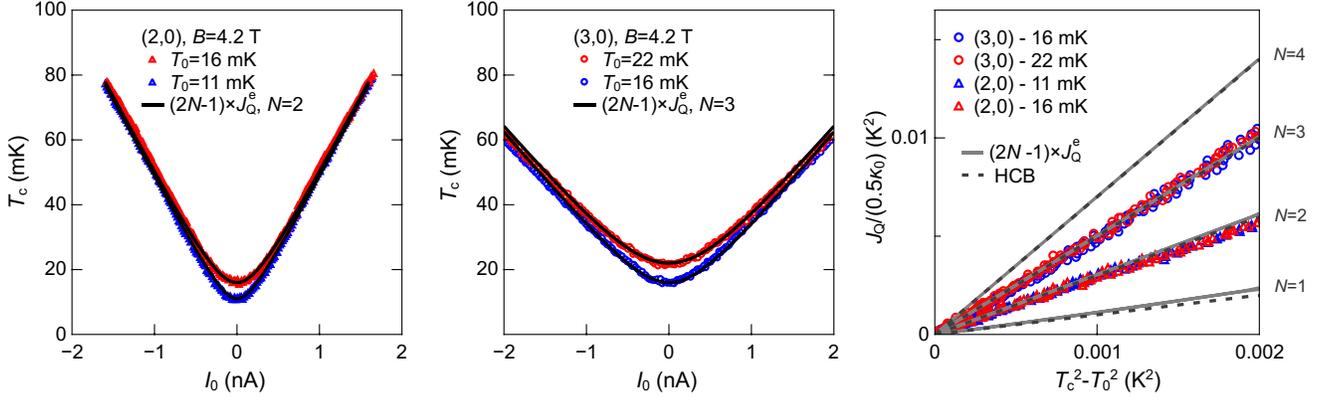

**Supplementary Figure 24** | Left: measured $T_c$ versus $I_0$ for the $(2, 0)$ state at $B = 4.2$ T (blue symbols: $T_0 = 11$ mK, red symbols: $T_0 = 16$ mK). Center: measured $T_c$ versus $I_0$ for the $(2, 0)$ state at $B = 4.2$ T (blue symbols: $T_0 = 16$ mK, red symbols: $T_0 = 22$ mK). Black lines: prediction of the heat balance model without HCB, with $2N - 1$ ballistic heat channels. Right: corresponding heat flow versus $\Delta(T_c^2)$. Full lines are prediction of the heat balance model without HCB, with $2N - 1$ ballistic heat channels. Dashed lines are predictions with HCB at $T_0 = 20$ mK.

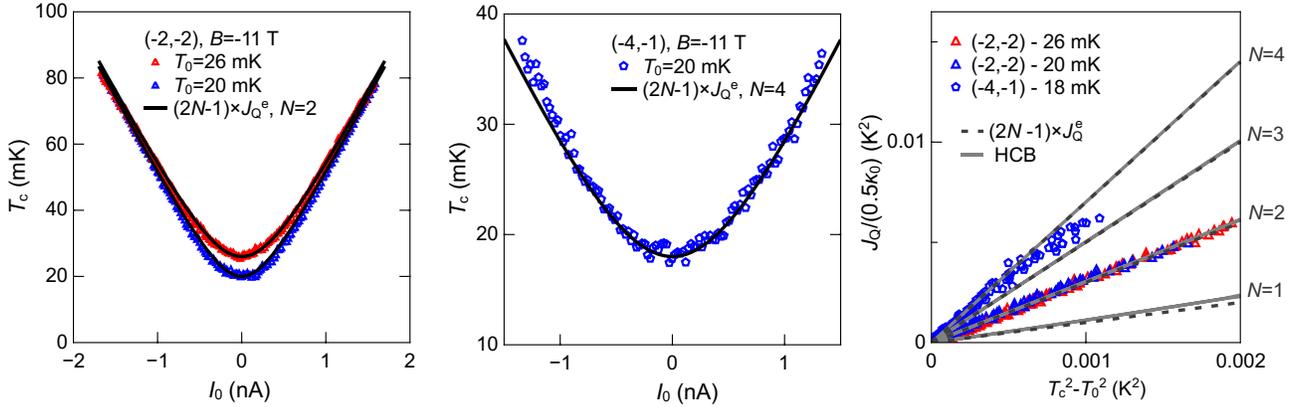

**Supplementary Figure 25** | Left: measured $T_c$ versus $I_0$ for the $(-2, -2)$ state at $B = -11$ T (blue symbols: $T_0 = 20$ mK, red symbols: $T_0 = 26$ mK). Center: measured $T_c$ versus $I_0$ for the $(-4, -1)$ state at $B = 4.2$ T (blue symbols: $T_0 = 20$ mK). Black lines: prediction of the heat balance model without HCB, with $2N - 1$ ballistic heat channels. Right: corresponding heat flow versus $\Delta(T_c^2)$. Full lines are prediction of the heat balance model without HCB, with $2N - 1$ ballistic heat channels. Dashed lines are predictions with HCB at $T_0 = 20$ mK.



\end{document}